\documentclass{revtex4-1}
\usepackage{latexsym, color, graphicx, comment}
\usepackage{amssymb}
\usepackage{amsmath}
\usepackage{hyperref}
\hypersetup{colorlinks=true,linkcolor=blue}

\newcommand{\vect}[1]{\mbox{\boldmath $#1$}}
\DeclareMathAlphabet{\mathbfsf}{\encodingdefault}{\sfdefault}{bx}{n}

\begin{document}

\title{Computing local sensitivity and tolerances for stellarator physics properties using shape gradients}

% repeat the \author .. \affiliation  etc. as needed
% \email, \thanks, \homepage, \altaffiliation all apply to the current author.
% Explanatory text should go in the []'s,
% actual e-mail address or url should go in the {}'s for \email and \homepage.
% Please use the appropriate macro for the type of information

% \affiliation command applies to all authors since the last \affiliation command.
% The \affiliation command should follow the other information.

\author{Matt Landreman}
\email[]{mattland@umd.edu}
\author{Elizabeth Paul}
%\homepage[]{Your web page}
%\thanks{}
\affiliation{Institute for Research in Electronics and Applied Physics, University of Maryland, College Park, MD, 20742, USA}

% Collaboration name, if desired (requires use of superscriptaddress option in \documentclass).
% \noaffiliation is required (may also be used with the \author command).
%\collaboration{}
%\noaffiliation


\date{\today}

\begin{abstract}

Tight tolerances have been a leading driver of cost in recent stellarator experiments,
so improved definition and control of tolerances can have significant impact on progress in the field.
Here we relate tolerances to the shape gradient representation that has been useful for shape optimization in industry,
used for example to determine which regions of
a car or aerofoil
most affect drag, and
we demonstrate how the shape gradient can be computed for physics
properties of toroidal plasmas. The shape gradient gives the local differential contribution to some scalar figure of merit (shape functional)
%(a functional of the shape) 
caused by 
normal displacement of the shape.
In contrast to derivatives with respect to quantities parameterizing a shape (e.g. Fourier amplitudes), 
which have been used previously for optimizing plasma and coil shapes,
the shape gradient gives spatially local information and so is 
more easily related to engineering constraints.
We present a method to determine 
the shape gradient for any figure of merit using the parameter derivatives
that are already routinely computed for stellarator optimization,
%The method is based on the solution of 
by solving a small linear system
relating shape parameter changes
to normal displacement.
Examples of shape gradients for plasma and electromagnetic coil shapes are given.
We also derive and present examples of an analogous representation of the local sensitivity to magnetic field errors;
this magnetic sensitivity can be rapidly computed from the shape gradient.
The shape gradient and magnetic sensitivity can both be converted into local tolerances, which inform
how accurately the coils should be built and positioned, where trim coils and structural supports for coils should be placed, and
where magnetic material and current leads can best be located.
Both sensitivity measures provide insight into shape optimization, 
enable systematic calculation of tolerances, and connect physics optimization to engineering criteria that are more easily specified in real space than in Fourier space.

\end{abstract}

% insert suggested PACS numbers in braces on next line

\maketitle %\maketitle must follow title, authors, abstract and \pacs

\section{Introduction}

Optimization of shapes -- both the shapes of plasmas and of electromagnetic coils -- is central to the modern stellarator
fusion concept. In the limit of low plasma $\beta$ (= plasma pressure $/$ magnetic pressure), the shape of an outer flux surface completely determines the magnetic field inside
up to an overall scale factor, and so this boundary shape represents a primary channel for controlling the confinement physics.
The shapes of electromagnetic coils are equally significant, for the design of a stellarator ultimately comes down to the selection of these coil shapes.
Moreover, \emph{derivatives} of various quantities with respect to the plasma shape and coil shape
are critical, for several reasons. These derivatives are used by gradient-based optimization algorithms for design,
and these derivatives also encode the tolerances to which the device must be built, tolerances which are a significant driver
for the cost of experiments (\cite{Strykowsky,Neilson}, `Lesson 1' in \cite{Klinger}). 
Indeed, an analysis of the cost overruns for the NCSX stellarator concluded 
`The largest driver of the project cost growth were the accuracy requirements' \cite{Strykowsky}.
Derivatives with respect to coil shape also are informative for determining how rigid the coils
must be and for designing the coils' support structure, so deformations of the coils under various loads do not detract from plasma performance.

There are a number of approaches to represent derivatives with respect to shape.
The  approach that has been used nearly exclusively to date for stellarators is to represent sensitivity using
the derivatives $\partial f/\partial p_j$,
where $f$ is some scalar figure of merit (i.e. a functional of the shape) like the rotational transform or neoclassical confinement on a particular flux surface,
and $p_j$ is a set of numbers that parameterize the shape, often Fourier amplitudes.
 We will call this first method the `parameterization approach'. However, in other fields such as aircraft and automotive design, a different representation of shape
sensitivity -- based on integrals of a `shape gradient' over surfaces -- has proved to be useful.
This second method, which we will call the `shape gradient approach',
has been described in several references \cite{Pironneau, Haug,SokolowskiZolesio, MohammadiBook, Haslinger, MohammadiPaper, ChoiKim, Dogan, Othmer2008, DelfourZolesio, Othmer, Walker}, and it has 
a number of advantages
over the parameterization approach. First, in contrast to the parameterization approach, the shape gradient approach provides 
sensitivity information that is spatially local. This difference is important because engineering constraints and tolerances are typically  specified in real space
rather than in Fourier space.
Moreover, spatially local sensitivity information can inform how and where coils are connected to their support structure,
so motion of the coils is minimized in high sensitivity regions.
Second, in the parameterization approach, there is typically a degeneracy in the sense that 
certain changes to the parameters move the surface tangent to itself and hence leave the shape unchanged; 
this degeneracy is absent in the  shape gradient approach.
Third, in contrast to the parameterization approach, the shape gradient approach provides coordinate-independent information,
so one can be less concerned with whether the parameterization chosen is an optimal one.
Motivated by these advantages, our goal in this paper is to introduce the  shape gradient representation to the stellarator community,
and to demonstrate an algorithm by which shape gradients can be computed using existing physics codes.

To more precisely define and contrast the parameterization and shape gradient approaches, 
we begin with the former, and consider a particular figure of merit $f$, the rotational transform on the magnetic axis, $\iota_0$.
In the widely used VMEC MHD equilibrium code \cite{VMEC1983, VMEC1986},
the plasma boundary shape is parameterized by the quantities $\left\{p_j\right\} = \left\{R_{m,n}^s,\, R_{m,n}^c,\,Z_{m,n}^s,\,Z_{m,n}^c\right\}$, 
which define the shape via
\begin{align}
R(\theta,\zeta) &= \sum_{m,n} \left[ R_{m,n}^s \sin(m \theta - n \zeta) + R_{m,n}^c \cos(m \theta - n \zeta) \right], \\
Z(\theta,\zeta) &= \sum_{m,n} \left[ Z_{m,n}^s \sin(m \theta - n \zeta) + Z_{m,n}^c \cos(m \theta - n \zeta) \right]. \nonumber
\end{align}
Here, $(R,\zeta,Z)$ are standard cylindrical coordinates, and $\theta$ is any poloidal angle. 
Thus, in the parameterization approach,  
the variation of $\iota_0$ with plasma shape is represented
by the quantities $\partial \iota_0/\partial R_{m,n}^c$, $\partial \iota_0/\partial Z_{m,n}^s$, etc.
(Other possible choices for $p_j$ exist, such as the Garabedian $\Delta_{m,n}$ coefficients \cite{Garabedian}.)
Similarly, coil shapes have been parameterized  \cite{focus} using
$\left\{p_j\right\} = \left\{X_{k,m}^s,\, X_{k,m}^c,\,Y_{k,m}^s,\, Y_{k,m}^c,\,Z_{k,m}^s,\, Z_{k,m}^c\right\}$, where
\begin{align}
\label{eq:coil_Fourier}
X_k(\vartheta) &= X_{k,0}^c + \sum_{m=1}^{m_{\max}} \left[ X_{k,m}^c \cos(m\vartheta) + X_{k,m}^s \sin(m\vartheta)\right] ,
\end{align}
with analogous expressions for $X\to Y$ and $X\to Z$.
Here, $(X,Y,Z)$ are Cartesian coordinates, $k$ indexes the various coils, and $\vartheta \in [0,2\pi)$. 
 (Here and throughout, we approximate coils as infinitesimally thin curves.)
Then in the parameterization approach,
variation of $\iota_0$ with coil shape is represented
by the quantities $\partial \iota_0/\partial X_{k,m}^c$, $\partial \iota_0/\partial X_{k,m}^s$, etc.
While these parameter derivatives have been used successfully for gradient-based optimization 
of plasma  \cite{stellopt_Spong,stellopt_Reiman}
and coil \cite{focus} shapes,
this representation of sensitivity does have the aforementioned downsides.
For instance, $\partial \iota_0/\partial R_{m,n}^c$, $\partial \iota_0/\partial X_{k,m}^c$, etc.~do not directly give local
information about which part of the plasma or coil  $\iota_0$ is most sensitive to.
It is not clear how to convert these derivatives into a local tolerance, 
or how to display these derivatives in a three-dimensional representation of the plasma or coil shape.
Also, $\partial \iota_0/\partial R_{m,n}^c$, $\partial \iota_0/\partial X_{k,m}^c$, etc. are not unique,
as they depend on the arbitrary choice of how $\theta$ and $\vartheta$ are defined.

\begin{figure}[h!]
\includegraphics[width=6.5in]{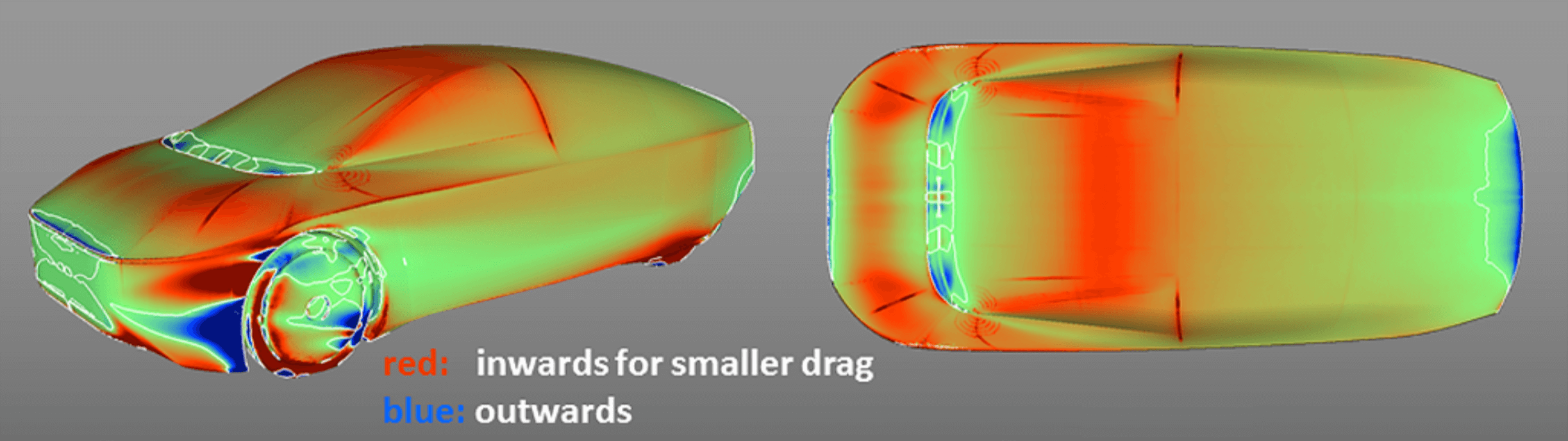}
\caption{(Color online)
Shape gradient $S$  for the figure of merit $f=$ total drag,
computed by researchers at Volkswagen for shape optimization of their vehicles.
Figure reproduced 
from \cite{Othmer}, licensed under \href{https://creativecommons.org/licenses/by/2.0/}{CC BY 2.0}; the original has been cropped.
\label{fig:car}}
\end{figure}

These shortcomings are absent in the shape gradient approach, which we can now define precisely.
Let $f$ be any scalar figure of merit that depends implicitly on the shape of the plasma boundary surface or on the shapes of the coils.
We then imagine the boundary surface or coil shapes are changed by a small amount $\delta \vect{r}$,
causing a perturbation $\delta f$ to $f$.
For shape functionals of the plasma boundary,
$\delta f$ is expressed in the form
\begin{equation}
\delta f = \int d^2a \; S \, \delta\vect{r} \cdot \vect{n},
\label{eq:Sdef3D}
\end{equation}
where $d^2a$ is an area integral, $\vect{n}$ is the unit vector normal to the surface, and $S$ is the shape gradient.
(While the term shape gradient is used by some authors \cite{Dogan, Walker, Dekeyser2014JCP, DekeyserThesis,Baelmans}, 
other authors use the name `density gradient' \cite{SokolowskiZolesio,DelfourZolesio} or `sensitivity map' \cite{Othmer2008,Othmer} for $S$,
while others do not give it a name \cite{Pironneau,Haug,ChoiKim}.)
For shape functionals of the coils, $\delta f$ is expressed in a similar way:
\begin{equation}
\delta f = \sum_{k } \int d\ell \; \vect{S}_k \cdot \, \delta\vect{r},
\label{eq:Sdef_3Dcurve}
\end{equation}
where $\ell$ is the arclength along a coil, $k$ indexes the coils, the shape gradient $\vect{S}_k$ is now a vector, and $\vect{S}_k$ has vanishing component tangent  to the curve. 
In both the surface and coil cases, the shape gradient represents the contribution to a differential change in the  figure of merit
due to local normal displacement of the shape.
The fact that perturbations can be expressed in the forms (\ref{eq:Sdef3D})-(\ref{eq:Sdef_3Dcurve}) for many $f$ 
has been addressed with great rigor elsewhere \cite{SokolowskiZolesio,DelfourZolesio};
we will motivate these expressions in section \ref{sec:existence} and provide a practical test for these expressions' validity in section \ref{sec:finite_difference}.
The shape gradients $S$ and $\vect{S}_k$ provide spatially local information, and they are independent of the particular
parameterization chosen for the shapes.
As shown in figure \ref{fig:car}, the shape gradient provides a very illuminating visualization of which parts
of a shape are critical for determining $f$.
This insight provided by the shape gradient is valuable for  human designers as they interact with optimization codes, as described for automotive shape design on page 11 of \cite{Othmer}.
Furthermore, the inverse of the magnitude of the shape gradient provides a local tolerance, in a sense that will be made precise in section \ref{sec:tolerance}.
Importantly, the forms (\ref{eq:Sdef3D})-(\ref{eq:Sdef_3Dcurve}) imply that a shape error of given $|\delta\vect{r}|$ has an effect
scaling with the area or length of the perturbation, so the tolerance for global shape errors is smaller than
the tolerance for localized errors.

If the representation (\ref{eq:Sdef3D}) exists (and if the boundary rotational transform is irrational),
we will  show that an expression similar to (\ref{eq:Sdef3D}) can be written in terms of the magnetic field perturbations
on the unperturbed plasma boundary:
\begin{equation}
\delta f =  \left< S\right> \delta V + \int d^2a \; S_B \, \delta\vect{B} \cdot \vect{n},
\label{eq:SB}
\end{equation}
where $\langle \ldots \rangle$ denotes a flux surface average, and $\delta V$ is an optional perturbation to the plasma volume
that may accompany the magnetic perturbation.
We will call $S_B$ the magnetic sensitivity.
Regions where $S_B$ is large are good locations for coils that control the relevant $f$ while having lesser effects
on other plasma properties.
Similarly to the case of shape perturbations, (\ref{eq:SB}) indicates that the tolerance for magnetic field errors
scales inversely with the surface area over which the error occurs.

The main ideas of this paper can now be summarized as follows. First, 
the shape gradients $S$ and $\vect{S}_k$ exist for  figures of merit $f$
that are of interest for magnetic confinement, as we will show in figures \ref{fig:sensitivity_iota_Fourier} and \ref{fig:sensitivity_NEO_Fourier}.
For any $f$,
the shape gradient (if it exists) can be computed from the derivatives of $f$
with respect to any shape parameters of the coils or boundary surface,
which are quantities already routinely computed for stellarator optimization.
This calculation of the shape gradient (section \ref{sec:finite_difference}) is done by recognizing equations (\ref{eq:Sdef3D})-(\ref{eq:Sdef_3Dcurve})
as linear integral equations that can be solved for $S$ or $\vect{S}_k$. 
Upon discretization, the resulting linear system also can be used to verify whether the shape gradient representation exists for a given $f$.
For instance, in the case of surfaces, the linear system is generally overdetermined, 
and one can check whether the derivatives of $f$ with respect to shape parameters lie in the column space
of the matrix representation of the integral operator. If a shape gradient for the bounding
toroidal surface (\ref{eq:Sdef3D}) is determined, the magnetic sensitivity $S_B$ can be
determined by solving a magnetic differential equation, (\ref{eq:SB_def}), as shown in figure \ref{fig:magnetic}.
Finally, the appropriately scaled inverses of $|\vect{S}_k|$ and $|S_B|$ give spatially local tolerances
on errors to coil shape and magnetic field, (\ref{eq:tolerance}) and (\ref{eq:magnetic_tolerance}),
as illustrated in figures \ref{fig:sensitivity_iota_coils} and \ref{fig:sensitivity_NEO_coils}.

Previous work on the shape gradient representation, for example in neutral fluid flow problems \cite{Othmer2008, YangStadler, Othmer},
has been closely associated with adjoint methods.
In that work, analytic manipulation of the relevant partial differential equation 
reveals a formula for the shape gradient in terms of the solution of an adjoint equation.
While adjoint methods are a very efficient method for computing the shape gradient, ideally
requiring only the cost of about one additional forward solve,
adjoint methods require non-negligible analytic
work and code development, and we will not explore adjoint methods further here. Instead, we will
develop procedures for computing the shape gradient from any existing `forward' code.

While the shape gradient representation has been used for some time in the realms of neutral fluids
and structural mechanics, the first application to fusion only recently appeared,
involving shape optimization for a tokamak divertor \cite{Dekeyser2012, Dekeyser2014NF, Dekeyser2014JCP, DekeyserThesis, Baelmans}.
To our knowledge, the  shape gradient representation has not been applied previously for stellarator plasma or coil shapes,
aside from our accompanying paper \cite{Elizabeth}.
In \cite{MynickPomphrey} it was pointed out that the redundancy in the parameterization method associated with tangential
displacements can be eliminated by considering only normal displacements, although the shape gradient representation was not displayed.
A magnetic sensitivity similar to $S_B$ 
was discussed in \cite{bnpert,BoozerKu2011}, although in that case the magnetic sensitivity
was computed by a different method to the one here, and shape gradients were not discussed.
Previous computations of tolerances on coil shapes  \cite{Yamazaki, Kremer, Williamson, Pedersen2006FST, Pedersen2006PoP, Andreeva, Bosch, Hartwell} 
have mostly been based on the size of magnetic islands
as computed from vacuum Poincare plots; little attention has been paid to the effect of shape and field errors on other physics figures of merit,
which we will consider here. With the exception of \cite{Williamson},
earlier coil tolerance calculations have typically been done by perturbing the Fourier amplitudes of the coil shapes,
yielding Fourier-space sensitivity information rather than the spatially local sensitivity we will compute.
The CNT device \cite{Kremer, Pedersen2006FST, Pedersen2006PoP} demonstrated that stellarators can be optimized to have generous tolerances,
and the tolerance expressions we compute here could be included into the objective functions in future `risk-averse' or `robust' optimizations.

Once the shape gradient is computed, it can be used for gradient-based minimization of $f$: the surface boundary 
is displaced by $-\epsilon S \vect{n}$, or the coil shape is displaced by $-\epsilon \vect{S}_k$, where $\epsilon>0$ is a step size 
that could be determined by a line search. (The shape Hessian can also be computed for use with optimization algorithms
that exploit second  derivatives \cite{ChoiKim, YangStadler}.) The shape gradient typically has lower regularity than the shape itself \cite{Othmer,Baelmans},
so smoothing is often applied to the gradient before the shape update. 
In practice, the overall objective function for optimization, $f_{total}$, is always a weighted sum of terms representing multiple criteria, $f_j$.
While the shape gradient for $f_{total}$ vanishes at an optimum, the shape gradient for any of the components $f_j$ generally do not.
Hence, second derivative information (which we will not consider here) is needed to understand variation in $f_{total}$ about
an optimum \cite{CaoxiangSensitivity}, but second derivative information is not necessary to understand the sensitivity of any individual criterion $f_j$ at the
optimum.

\section{Existence of a shape gradient}
\label{sec:existence}

Let us next motivate that the shape gradient representations (\ref{eq:Sdef3D})-(\ref{eq:Sdef_3Dcurve}) can be expected to exist for a 
large class of shape functionals $f$. 
The central point is that (\ref{eq:Sdef3D})-(\ref{eq:Sdef_3Dcurve}) are essentially the generalization of the standard chain rule to
functions that depend on an infinite number of parameters, i.e. functionals.
Consider a simpler 1D problem in which a scalar function $f$ depends on a discrete number of parameters:
$f=f(r_1, r_2, \ldots, r_n)$. When the $r_j$ parameters are perturbed, $r_j \to r_j + \delta r_j$, the function varies $f \to f + \delta f$
according to
\begin{equation}
\delta f = \sum_{j=1}^n \frac{\partial f}{\partial r_j} \delta r_j .
\label{eq:total_variation}
\end{equation}
In the limit $n \to \infty$, the $r_j$ are replaced with a function $r(\vartheta)$,
$f$ becomes a functional of $r(\vartheta)$,
the sum in (\ref{eq:total_variation}) becomes an integral,
and the finite set of numbers $\partial f/\partial r_j$ is replaced by some $\vartheta$-dependent function which can be denoted $\delta f/\delta  r$:
\begin{equation}
\delta f = \int_0^{2\pi} d\vartheta \, \frac{\delta f}{\delta r} \, \delta r.
\label{eq:functional_derivative}
\end{equation}
Here, $ \delta f/\delta r$ is the `functional derivative' arising in the calculus of variations.
The fact that $\delta f$ can be written in the form (\ref{eq:functional_derivative}) is also \cite{Dogan, MorinNochetto} an instance of the Riesz representation theorem,
which (roughly) states that any linear operator can be written as an inner product with some element of the appropriate space.
In our case, $\delta f$ is 
a linear operator acting on $\delta r$, the inner product is integration over $\vartheta$, and the `element of the appropriate space' is $\delta f/\delta r$.
The quantity $\delta f/\delta r$ in (\ref{eq:functional_derivative}) will become -- after straightforward generalization to multiple dimensions and rescaling -- the shape gradient.

Generalizing (\ref{eq:functional_derivative}) to the case in which $f$ depends on three functions $r_X(\vartheta)$, 
$r_Y(\vartheta)$, and $r_Z(\vartheta)$,
\begin{equation}
\delta f = \int_0^{2\pi} d\vartheta \frac{\delta f}{\delta r_X} \delta r_X + \int_0^{2\pi} d\vartheta \frac{\delta f}{\delta r_Y} \delta r_Y + \int_0^{2\pi} d\vartheta \frac{\delta f}{\delta r_Z} \delta r_Z.
\label{eq:3D_variation}
\end{equation}
Defining the position vector
$\vect{r} = \vect{e}_X r_X + \vect{e}_Y r_Y +\vect{e}_Z r_Z$
and
$\vect{S} = | d\vect{r}/d\vartheta|^{-1}( \vect{e}_X \delta f/\delta r_X + \vect{e}_Y \delta f/\delta r_Y + \vect{e}_Z \delta f/\delta r_Z)$
where $\vect{e}_{X,Y,Z}$ are Cartesian unit vectors, then 
we obtain
the single-coil case of (\ref{eq:Sdef_3Dcurve}): $\delta f = \int d\ell \; \vect{S} \cdot \delta \vect{r}$.
If $f$ depends only on the shape of the curve and not on its particular parameterization,
then a perturbation $\delta\vect{r}$ parallel to $d\vect{r}/d\ell$ must cause no $\delta f$, so $\vect{S}$ has no tangential component:
$\vect{S} \cdot d\vect{r}/d\ell=0$. 
Eq (\ref{eq:Sdef_3Dcurve}) is a straightforward generalization
to multiple curves. Note that for curves confined to a plane, a scalar shape gradient $S = \vect{S}\cdot\vect{n}$ can be defined to obtain
\begin{equation}
\delta f = \int d\ell \; S \, \delta\vect{r} \cdot \vect{n}.
\label{eq:Sdef2D}
\end{equation}

The above analysis is extended to surfaces if we let the $r_{X,Y,Z}$ depend on two independent variables $(\theta,\zeta)$. Then (\ref{eq:3D_variation})
becomes
\begin{equation}
\delta f = \int_0^{2\pi} d\theta \int_0^{2\pi} d\zeta \frac{\delta f}{\delta r_X} \delta r_X + \int_0^{2\pi} d\theta \int_0^{2\pi} d\zeta \frac{\delta f}{\delta r_Y} \delta r_Y + \int_0^{2\pi} d\theta \int_0^{2\pi} d\zeta \frac{\delta f}{\delta r_Z} \delta r_Z.
\end{equation}
Now defining 
$\vect{S} = N^{-1}( \vect{e}_X \delta f/\delta r_X + \vect{e}_Y \delta f/\delta r_Y + \vect{e}_Z \delta f/\delta r_Z)$
where $N=|\vect{N}|$ and $\vect{N} = \partial \vect{r}/\partial\zeta \times \partial\vect{r}/\partial\theta$, we obtain
$\delta f = \int d^2a \; \vect{S} \cdot \delta \vect{r}$.
If $f$ depends only on the shape of the surface and not on its particular parameterization,
then a perturbation $\delta\vect{r}$ parallel to any tangent vector must cause no $\delta f$.
Therefore $\vect{S}$ has only a normal component: $\vect{S} = \vect{n} S$, yielding (\ref{eq:Sdef3D}).

Note that the shape gradient representation (\ref{eq:Sdef3D}) does not exist for any figure of merit $f$ that depends on the specific coordinates
used to parameterize the surface (if $f$ cannot be expressed in a coordinate-independent way). For such an $f$, then $\delta f$ would depend also
on the components of $\delta \vect{r}$ tangent to the surface, not only the normal component. Similarly, in the representation 
(\ref{eq:Sdef_3Dcurve}) for such an $f$, $\vect{S}_k$ would have a component tangent to the curve.
If $f$ depends on quantities other than the shape, such as the radial profiles of plasma pressure and current, 
these other variables can  be understood to  be implicit functions of the shape;
or, these quantities could be considered to be additional independent variables, giving rise to additional terms in the expressions for $\delta f$.

The conditions for existence of the shape gradient representation have been analyzed
with greater mathematical rigor in section 2.11 of \cite{SokolowskiZolesio} and section 9.3.4 of \cite{DelfourZolesio}.
The literature contains analytic calculations of the shape gradient for shape functionals in several problems,
such as variants of Poisson's equation \cite{Pironneau}, the Navier-Stokes equations
\cite{Pironneau, Othmer2008, YangStadler},  structural mechanics problems \cite{Haug,ChoiKim},
and plasma fluid equations \cite{Dekeyser2014JCP}.
It is beyond the scope of this paper to give rigorous explicit proofs that the derivatives of  figures of merit arising in stellarator optimization
can be represented in the shape gradient form, but in the next section we will give derivations for a few examples.
Also, in section \ref{sec:finite_difference} we will demonstrate
a numerical method that can test whether any particular figure of merit can or cannot be represented this way.

\section{Examples}
\label{sec:examples}

Let us now show explicitly that for certain figures of merit (shape functionals) $f$, 
changes caused by perturbations to curves or surfaces
can indeed be expressed in the forms (\ref{eq:Sdef3D})-(\ref{eq:Sdef_3Dcurve}). 
These examples are valuable since they will be used to verify the algorithms in section \ref{sec:finite_difference},
before the algorithms are applied to more complicated functionals for which analytic expressions for the shape gradient are unavailable.

\subsection{Volume integrals}
\label{sec:volume_integrals}

Perhaps the simplest example is $f=$ the volume enclosed by a surface.
The change in volume associated with a perturbation $\delta\vect{r}$ to the boundary is a sum over the boundary of differential volume elements.
Each differential volume element has an area $d^2a$ along the unperturbed surface and height $\vect{n}\cdot\delta \vect{r}$ perpendicular to it.
Summing the volume of these elements, we see $\delta f$ has the form (\ref{eq:Sdef3D}) with shape gradient $S=1$. 
More generally, by the same geometric reasoning,
for the volume integral $f=\int d^3r\, Q(\vect{r})$ of any quantity $Q(\vect{r})$,
 $\delta f$ has the form (\ref{eq:Sdef3D}) with shape gradient $S=Q$. 

\subsection{Integrals along a curve}
\label{sec:curve_integrals}

Consider the integral of any quantity $Q(\vect{r})$ along a closed curve parameterized by $\vartheta$:
\begin{equation}
L = \int Q\, d\ell = \int_0^{2\pi}d\vartheta \; Q\left| \frac{d\vect{r}}{d\vartheta} \right|.
\end{equation}
Perturbing the curve shape $\vect{r} \to \vect{r} + \delta\vect{r}$,
the associated change to the integral is
\begin{equation}
\delta L = \int_0^{2\pi}d\vartheta \left(
\left| \frac{d\vect{r}}{d\vartheta} \right| \delta\vect{r}\cdot\nabla Q
+
Q\left| \frac{d\vect{r}}{d\vartheta} \right|^{-1}
\frac{d\vect{r}}{d\vartheta} \cdot 
\frac{d \delta \vect{r}}{d\vartheta}
 \right).
\end{equation}
Integrating by parts to remove the $\vartheta$ derivative from $\delta \vect{r}$ in the last term,
\begin{equation}
\delta L = \int d\ell \, \delta\vect{r} \cdot \left[ (\vect{I}-\vect{t}\vect{t})\cdot\nabla Q - Q \kappa \vect{n}\right],
\label{eq:space_curve}
\end{equation}
where $\vect{I}$ is the identity tensor and $\vect{t} = d\vect{r}/d\ell = |d\vect{r}/d\vartheta|^{-1} d\vect{r}/d\vartheta$ is the unit tangent vector.
Here, the curvature $\kappa$ (equal in magnitude to the inverse radius of curvature)
and normal vector $\vect{n}$ are 
defined by $\kappa \vect{n} = d \vect{t}/d\ell = |d\vect{r}/d\vartheta|^{-2} (\vect{I}-\vect{t}\vect{t})\cdot d^2\vect{r}/d\vartheta^2$. Thus, perturbations to the shape
indeed cause $L$ to vary according to the form (\ref{eq:Sdef_3Dcurve}) (with a single term in the sum) with shape gradient 
\begin{equation}
\vect{S} = (\vect{I}-\vect{t}\vect{t})\cdot\nabla Q - Q \kappa \vect{n}.
\label{eq:S_curve}
\end{equation}
As required,
$\vect{S}$ has vanishing component tangent to the curve.
In the case of a plane curve, we can make the substitution $\vect{I} = \vect{t}\vect{t}+\vect{n}\vect{n}$, causing 
(\ref{eq:space_curve}) to reduce to (\ref{eq:Sdef2D}) with a scalar shape gradient $S = \vect{n}\cdot\nabla Q - Q \kappa$.

\begin{figure}[h!]
\includegraphics[width=2.0in]{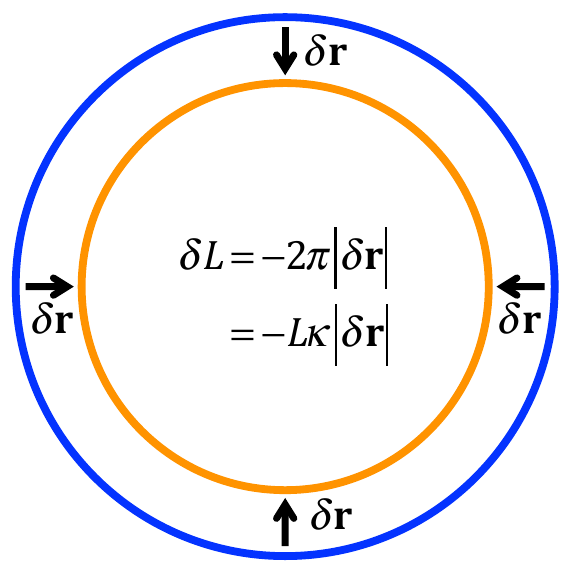}
\caption{(Color online)
Illustration of (\ref{eq:S_curve}): normal displacement of a curve in the direction $\kappa \vect{n}=d\vect{t}/d\ell$ causes the length to decrease.
\label{fig:curvature}}
\end{figure}

For both space curves and plane curves, if we make the choice $Q=1$, then $L$ becomes the length of the curve. The shape gradients then become $\vect{S} = -\kappa \vect{n}$ and $S = -\kappa$.
This result can be understood geometrically, as shown in figure \ref{fig:curvature}.
If a curve is given a normal displacement toward a center of curvature, the length decreases.

\subsection{Area integrals}
\label{sec:area}

Consider the area integral over a toroidal surface (defined by $\vect{r}(\theta,\zeta)$) of any quantity $Q(\vect{r})$:
\begin{equation}
A = \int d^2a\, Q = \int_0^{2\pi} d\theta \int_0^{2\pi} d\zeta \; N Q,
\label{eq:A}
\end{equation}
where $N=|\vect{N}|$, and
\begin{equation}
\vect{N} = \frac{\partial\vect{r}}{\partial\zeta} \times \frac{\partial\vect{r}}{\partial\theta}
=N \vect{n}
\label{eq:bigN}
\end{equation}
is a (non unit length) normal vector. As shown in appendix \ref{appendix:area}, by perturbing $\vect{r}(\theta,\zeta)$
in these expressions, we obtain the form (\ref{eq:Sdef3D}) with a shape gradient
\begin{equation}
S = \vect{n}\cdot\nabla Q - 2 Q H,
\label{eq:area_integrals}
\end{equation}
where $H$ is the mean curvature (defined in (\ref{eq:S_area_mean_curvature}).)
For the choice $Q=1$, corresponding to $A=$ the area of the surface, the shape gradient is just $-2H$.

\begin{figure}[h!]
\includegraphics[width=3.5in]{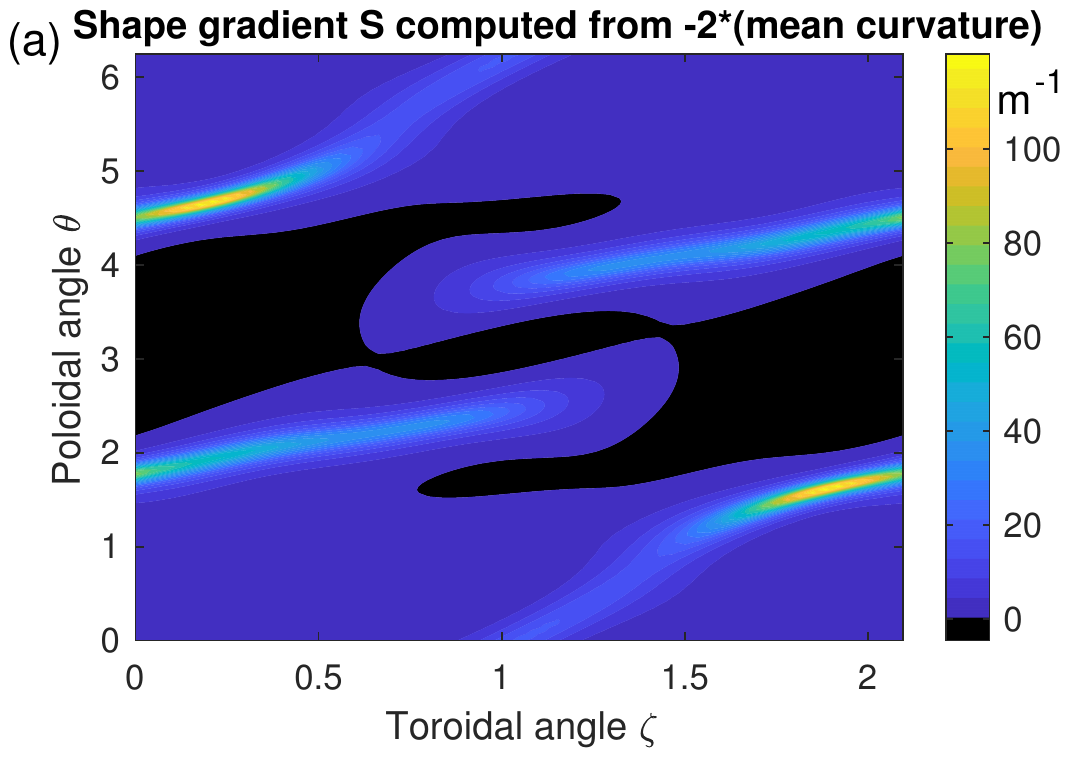}
\includegraphics[width=3.25in]{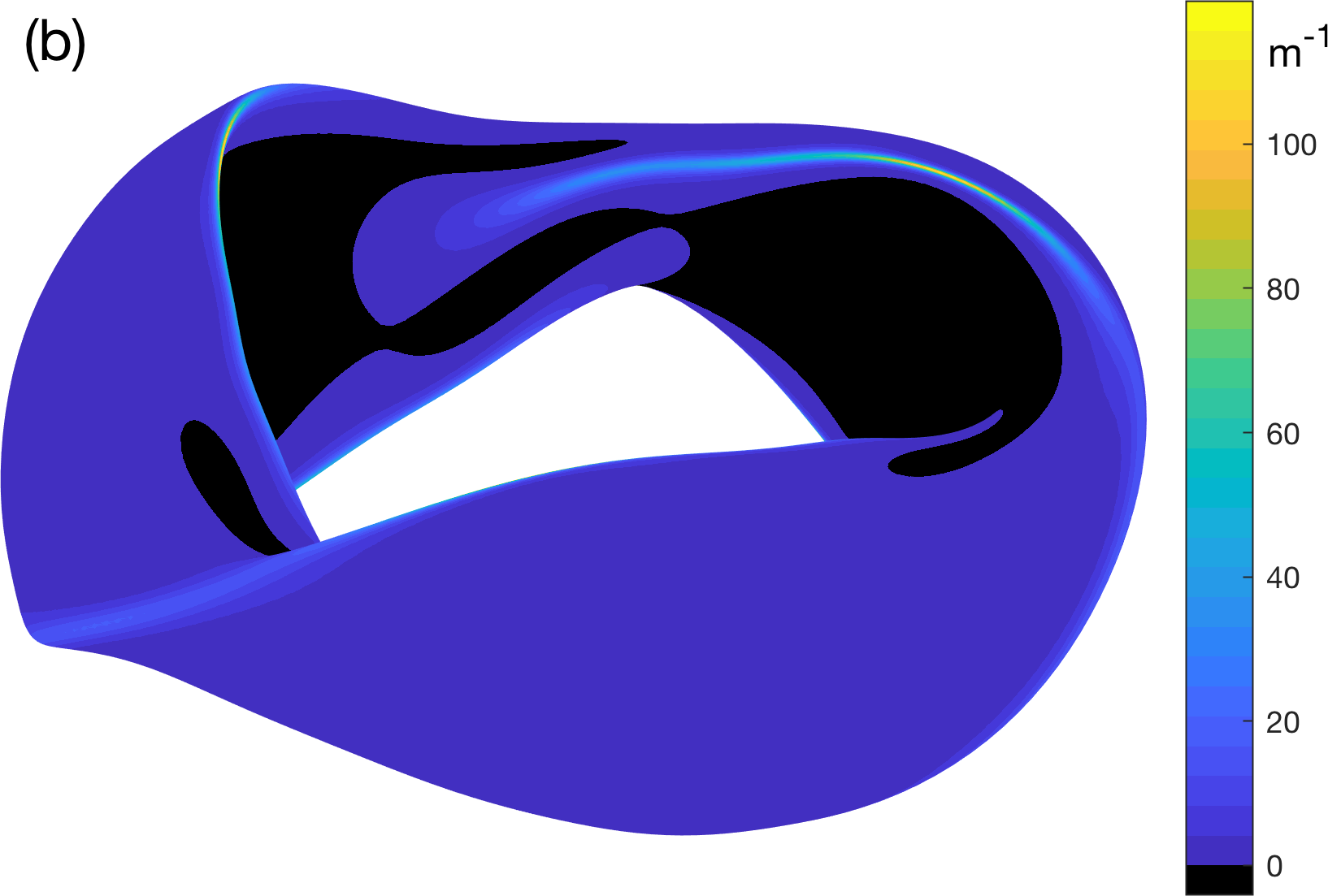}
\caption{(Color online)
Shape gradient for surface area in NCSX, computed using (\ref{eq:area_integrals}) and (\ref{eq:S_area_mean_curvature}).
\label{fig:sensitivity_area_3D}}
\end{figure}

This result can be interpreted geometrically, as in the earlier case of the 
curve (figure \ref{fig:curvature}).
If the toroidal surface is moved outward in a convex region, the area increases, whereas
the area decreases if the surface is moved outward in a concave region.
An example is shown in figure \ref{fig:sensitivity_area_3D} for the NCSX stellarator \cite{Zarnstorff}.
(The equilibrium LI383 is used.)
It can be seen that the shape gradient is highly localized to the sharp
edges of the plasma shape. The concave regions on the inboard side of the plasma
have a shape gradient that is slightly negative, drawn in black in the figures.

The results of this section can also be derived by a number of other methods, such as using the appropriate multidimensional Leibniz rule for differentiating integrals over moving domains. 
The shape gradient for area integrals (\ref{eq:area_integrals}) can be found as Lemma 1 on page 87 of \cite{Pironneau},
and as a simplification of eq (12) in \cite{DewarHudsonPrice}.
The above results for the shape gradient of volume and area integrals
can also be found as 
equations (6.38) and (6.53) in \cite{ChoiKim},
and equations (9.4.7) and (9.4.17) in \cite{DelfourZolesio}.

%%%%%%%%%%%%%%%%%%%%%%%%%%%%%%%%%%%%%%%%%%%%%%%%%%%%%%
%%%%%%%%%%%%%%%%%%%%%%%%%%%%%%%%%%%%%%%%%%%%%%%%%%%%%%
%%%%%%%%%%%%%%%%%%%%%%%%%%%%%%%%%%%%%%%%%%%%%%%%%%%%%%
%%%%%%%%%%%%%%%%%%%%%%%%%%%%%%%%%%%%%%%%%%%%%%%%%%%%%%

\section{Relation to derivatives with respect to shape parameters}
\label{sec:finite_difference}

In contrast to the examples in the previous section,
most interesting physics figures of merit $f$ are not obviously expressible as integrals over the coil shapes,
over the plasma volume, or over the plasma boundary surface.
Hence it may be impractical to analytically manipulate
the definitions of the figures of merit to reveal the shape gradient,
as we have done in the previous section.
Instead, we now show how the shape gradient -- if it exists -- can be computed from the derivatives 
that are routinely computed for stellarator optimization (i.e. the derivatives with respect to shape parameters.)
These derivatives are represented by the vector with elements $\partial f/\partial p_j$, denoted $\partial f/\partial\vect{p}$, 
where again $p_j$ denotes the numbers used to parameterize the shape. For the case of surfaces, $p_j$
could represent the $R_{m,n}^c$ and $Z_{m,n}^s$ Fourier coefficients, or $p_j$ could represent the Garabedian $\Delta_{m,n}$
coefficients. For the case of coils, $p_j$ could represent the Fourier coefficients of the Cartesian components of the coil shapes.
The method for computing the shape gradient from these parameter derivatives for coils is slightly different from the method for surfaces, 
and we will describe the method for coils first.

%%%%%%%%%%%%%%%%%%%%%%%%%%%%%%%%%%%%%%%%%%%%%%%%%%%%%%
%%%%%%%%%%%%%%%%%%%%%%%%%%%%%%%%%%%%%%%%%%%%%%%%%%%%%%

\subsection{Coils}
\label{sec:finite_differences_coils}

To compute the shape gradient for coils, we first note that an equivalent expression to (\ref{eq:Sdef_3Dcurve}) is
\begin{equation}
\int_0^{2\pi} d\vartheta \left| \frac{d\vect{r}}{d\vartheta}\right|  \frac{\partial\vect{r}}{\partial p_j} \cdot \vect{S}
= \frac{\partial f}{\partial p_j}
\label{eq:Sdef3Dcurves_alt}
\end{equation}
for all $j$. Here and throughout this section we suppress the coil number $k$ and $\sum_k$ to simplify notation. Our basic
method is to recognize (\ref{eq:Sdef3Dcurves_alt}) as a linear integral equation for $\vect{S}$. 
This equation can determine $\vect{S}$ to within some precision associated with the number of parameters $p_j$.
We can discretize $(S_X, \, S_Y,\, S_Z)$, the Cartesian components
of $\vect{S}$, in the same way as (\ref{eq:coil_Fourier}):
\begin{align}
\label{eq:S_coil_Fourier}
S_X(\vartheta) &= S_{X,0}^c + \sum_{m=1}^{m_{\max}} \left[ S_{X,m}^c \cos(m\vartheta) + S_{X,m}^s \sin(m\vartheta)\right], 
\end{align}
with analogous expressions for $S_X\to S_Y$ and $S_X\to S_Z$. If the same maximum mode number $m_{\max}$ is used for
the coil representation (\ref{eq:coil_Fourier}) as for (\ref{eq:S_coil_Fourier}),
then the number of discrete degrees of freedom in $\vect{S}$ is $6m_{\max}+3$, the same as the number of coil parameters $p_j$.
Thus, the linear system corresponding to the discretization of (\ref{eq:Sdef3Dcurves_alt}) is square.
As $m_{\max}$ is increased, $\vect{S}$ is recovered with greater accuracy, at the cost of having to evaluate
a larger number of parameter derivatives $\partial f/\partial p_j$.
Once $\vect{S}$ is recovered, we can check whether it has a component tangent to the curve. If the tangential
component does not converge towards zero with increasing resolution, then
$f$ is evidently not coordinate-independent and hence not physical.
While we have used a Fourier discretization here, the same approach can be applied to other discretizations such as splines.

\begin{figure}[h!]
\includegraphics[width=3.5in]{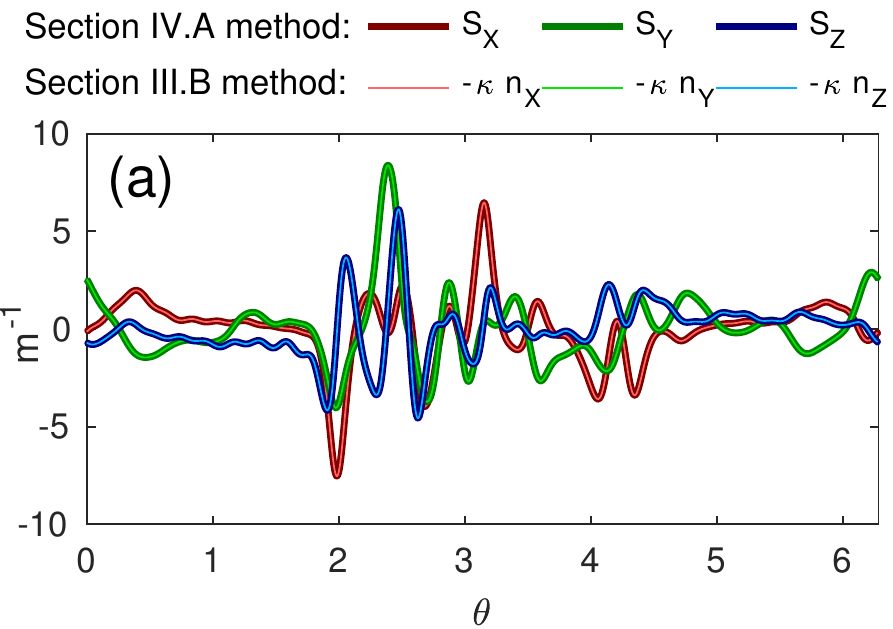}
\hspace{0.5in}
\includegraphics[height=2.5in]{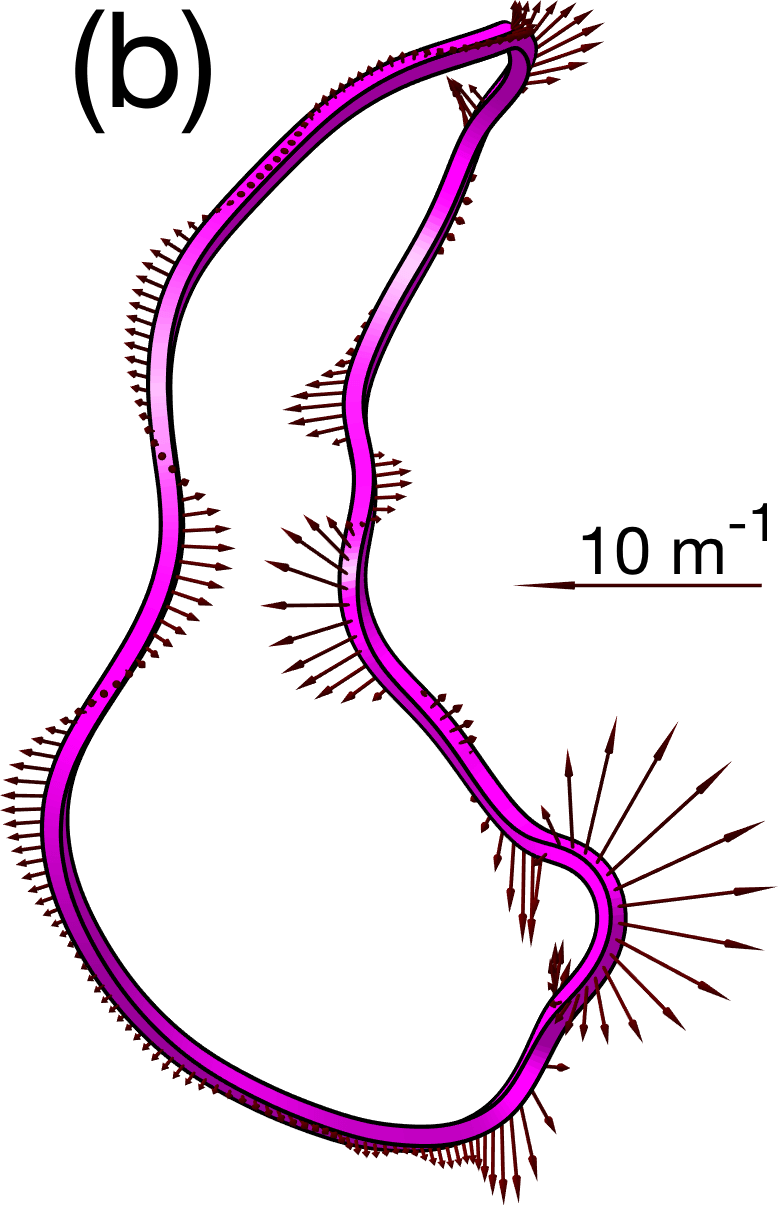}
\caption{(Color online)
(a) The shape gradient for length for the NCSX type-A modular coil, calculated two ways.
The thick curves are computed by evaluating the derivatives of length with respect to coil shape parameters ($\partial L/\partial p_j$)
using finite differences,
and then solving the system (\ref{eq:Sdef3Dcurves_alt}) for $\vect{S}$. The thin curves show the result expected from the analytic calculation of section
(\ref{sec:curve_integrals}).
The two methods yield indistinguishable results on the scale of the plot for each Cartesian component. 
(b) The same shape gradient displayed in 3D.
\label{fig:sensitivity_coil_length}}
\end{figure}

This procedure is demonstrated in figure \ref{fig:sensitivity_coil_length}, using the type-A modular coil from NCSX.
Here, the shape functional we consider is the 
curve length $L$, so from section (\ref{sec:curve_integrals}) we know analytically the true shape gradient is $-\kappa\vect{n}$.
To test the procedure described above, we evaluate the derivatives $\partial L/\partial X_m^s$, $\partial L/\partial X_m^c$, 
$\partial L/\partial Y_m^s$, etc.~using finite differences in the Fourier amplitudes. 
(For this $f$, the derivatives could be computed analytically, but the point here is to demonstrate a procedure that can be applied when only finite
difference derivatives are available.)
The system (\ref{eq:Sdef3Dcurves_alt})
is then solved for the Cartesian components of $\vect{S}$. As shown in figure  \ref{fig:sensitivity_coil_length}.a, the two
approaches give results that are indistinguishable on the scale of the plot, so we can have confidence applying the procedure to
other figures of merit for which the shape gradient is not available analytically. 
Figure \ref{fig:sensitivity_coil_length}.b
displays the shape gradient in three dimensions. 
Note that the scale for the shape gradient (arrow lengths) is independent of the scale for the coil itself.

%%%%%%%%%%%%%%%%%%%%%%%%%%%%%%%%%%%%%%%%%%%%%%%%%%%%%%
%%%%%%%%%%%%%%%%%%%%%%%%%%%%%%%%%%%%%%%%%%%%%%%%%%%%%%

\subsection{Surfaces}
\label{sec:finite_differences_surfaces}

The calculation of shape gradient $S$ from parameter derivatives for surfaces is a bit more complicated
than the calculation for coils due to the different number of dimensions involved. 
To compute the shape gradient for a surface, we first note that an equivalent expression to (\ref{eq:Sdef3D}) is
\begin{equation}
\int d^2a \; S \frac{\partial\vect{r}}{\partial p_j} \cdot \vect{n}
= \frac{\partial f}{\partial p_j}
\label{eq:Sdef3D_alt}
\end{equation}
for all $j$. Again we recognize (\ref{eq:Sdef3D_alt}) as a linear integral equation for $S$. Upon discretization,  (\ref{eq:Sdef3D_alt})
becomes a dense linear system. However this time the matrix for this linear system is generally not square, since 
the number of degrees of freedom used to discretize $S$ 
generally differs from the number of $p_j$. Therefore the right-hand side of 
(\ref{eq:Sdef3D_alt}) need not be in the column space of the matrix, associated with the fact that the shape gradient
representation does not exist for every $f$. One can therefore determine if a shape gradient exists for a given $f$ by checking whether the
right-hand side is in the column space of the matrix.

There are two variants of the method: $S$ can be discretized using values on grid points in $\theta$ and $\zeta$, or $S$ can be discretized in a finite Fourier expansion.
We call these two variants the `collocation method' and `Fourier method' respectively. (Note that even if $S$
is discretized using a Fourier expansion, it can still be evaluated at any point in $\theta$ and $\zeta$, and so the Fourier
method still results in spatially local sensitivity information.)
Based on experience so far with both methods, the Fourier method is somewhat more robust,
since the collocation method sometimes requires tuning of the number of singular values retained in a pseudoinverse,
whereas the Fourier method does not require any tuning.
Therefore, we advocate the Fourier method and focus on it here. The collocation method is discussed in appendix \ref{sec:collocation}.

In the Fourier method, we define the matrix $\vect{D}$, with matrix elements $D_{jq}$, by
\begin{equation}
D_{jq} 
= \int_0^{2\pi}d\theta \int_0^{2\pi}d\zeta \frac{\partial\vect{r}}{\partial p_j} \cdot \vect{N}  \exp\left( i m_q \theta - i n_q \zeta \right),
\label{eq:D_def}
\end{equation}
where a sequence of poloidal and toroidal mode numbers $(m,n)$ has been indexed by $q$,
and $\vect{N}$ is defined in (\ref{eq:bigN}).
Any poloidal and toroidal angles can be used, including angles in which the field lines are not straight; 
since the fundamental definition of the shape gradient (\ref{eq:Sdef3D}) is coordinate-independent,
our procedure will yield results independent of the choice of $\theta$ and $\zeta$ (up to discretization error).
The central idea of our method can now be stated as follows:

\emph{
A shape gradient for $f$ exists if and only if $\partial f/\partial\vect{p}$
lies in the column space of $\vect{D}$. If the shape gradient $S$ does exist, then it
can be obtained by solving}
\begin{equation}
\sum_q D_{jq} S_q = \frac{\partial f}{\partial p_j}	
\label{eq:main_result}
\end{equation}
\emph{ 
for the Fourier coefficients $S_q$ in}
\begin{equation}
S(\theta,\zeta) = \sum_q S_q \exp\left( i m_q \theta - i n_q \zeta\right).
\label{eq:S_Fourier}
\end{equation}
To justify this statement, 
suppose $\partial f/\partial\vect{p}$ lies in the column space of $\vect{D}$, so a solution to the linear system (\ref{eq:main_result}) for $S_q$ exists.
We can then form (\ref{eq:S_Fourier}) and evaluate
\begin{equation}
\int d^2a \; S \frac{\partial\vect{r}}{\partial p_j} \cdot \vect{n}
= \sum_q 
\int d^2a \; S_q \exp(i m_q \theta - i n_q \zeta) \frac{\partial\vect{r}}{\partial p_j} \cdot \vect{n} 
= \sum_q S_q  D_{jq}  = \frac{\partial f}{\partial p_j},
\label{eq:proof0}
\end{equation}
showing that $S$ in (\ref{eq:S_Fourier}) satisfies the defining property of a shape gradient, (\ref{eq:Sdef3D_alt}).
Conversely,
suppose the gradient $S$ exists. Then the Fourier coefficients $S_q$ exist, and (\ref{eq:S_Fourier}) can be substituted into 
(\ref{eq:Sdef3D_alt}), yielding (\ref{eq:main_result})
and implying that $\partial f/\partial\vect{p}$
lies in the column space of $\vect{D}$.

Stellarator shapes often posses `stellarator symmetry': $R(\theta,\zeta) = R(-\theta,-\zeta)$
and $Z(\theta,\zeta)=-Z(-\theta,-\zeta)$, implying $R_{m,n}^s=0$ and $Z_{m,n}^c=0$ for all $(m,n)$. In this case, (\ref{eq:D_def}) and (\ref{eq:S_Fourier}) can be replaced by
\begin{equation}
D_{jq} 
= \int_0^{2\pi}d\theta \int_0^{2\pi}d\zeta \frac{\partial\vect{r}}{\partial p_j} \cdot \vect{N}  \cos\left( m_q \theta - n_q \zeta \right)
\label{eq:D_def_sym}
\end{equation}
and
\begin{equation}
S(\theta,\zeta) = \sum_q S_q \cos\left(m_q \theta - n_q \zeta\right)
\label{eq:S_Fourier_sym}
\end{equation}
respectively; (\ref{eq:main_result}) and the rest of the reasoning above is unchanged. 
Stellarators shapes are also typically symmetric under toroidal rotation by $2\pi/n_{fp}$ for some integer $n_{fp}$ (e.g. $n_{fp}=3$ for NCSX.)
This `$n_{fp}$ symmetry' implies that only integer multiples of $n_{fp}$ need to be included in the $\{n_q\}$.
Several other technical points related to symmetry are discussed in appendix \ref{sec:symmetry}.

In the common (but not mandatory) situation that the same maximum $m$ is
used for $R_{m,n}^c$, $Z_{m,n}^s$, and $S_q$, and that the same is true of the maximum $n$ used,
then $\vect{D}$ will have approximately twice as many rows as columns. The ratio is not exactly 2 since $R_{m,n}^c$ and $S_q$
include a $m=n=0$ contribution but $Z_{mn}^s$ does not.
The fact that $\vect{D}$ has more rows than columns (i.e. the system is over-constrained) reflects the fact that
a shape gradient does not exist for every $f$ (in particular, coordinate-dependent $f$ which vary under tangential displacement.)

To make practical use of the results above, we can use either the $QR$ decomposition or singular value decomposition (SVD) of $\vect{D}$.
The $QR$ decomposition is $\vect{D}=\vect{Q}\vect{R}$ where
$\vect{Q}$ is a square orthogonal matrix ($\vect{Q}^{-1} = \vect{Q}^T$), and $\vect{R}$ is upper-triangular with the same dimensions as $\vect{D}$ (generally rectangular.) 
Here we are interested in the case in which $\vect{D}$ has $P$ rows and $M$ columns with $P>M$.
It follows that $\partial f/\partial\vect{p}$ is in the column space of $\vect{D}$ if and only if the vector $\vect{Q}^T \, \partial f/\partial \vect{p}$ is zero
for rows $>M$ and for any rows in which $\vect{R}$ has a nonzero diagonal entry. 
(For this problem, $\vect{D}$ generally has full rank, so all the diagonal entries of $\vect{R}$ are nonzero.)
If $\partial f/\partial\vect{p}$ is in the column space of $\vect{D}$, the solution to (\ref{eq:main_result}) is then given by $\vect{S}=\vect{R}_1^{-1} \vect{Q}_1^T\, \partial f/\partial \vect{p}$ where
where $\vect{S}$ is the vector with elements $S_q$, $\vect{R}_1$ is the $M\times M$
upper sub-matrix of $\vect{R}$, and $\vect{Q}_1$ is the $P\times M$ sub-matrix of $\vect{Q}$.
(The matrix $\vect{R}_1^{-1} \vect{Q}_1^T$ which is applied to $\partial f/\partial\vect{p}$ is the pseudo-inverse of $\vect{D}$ in this case.)
The SVD approach is similar: the SVD is
\begin{equation}
\vect{D} = \vect{U} \vect{\Sigma} \vect{V}^T,
\end{equation}
where $\vect{U}$ and $\vect{V}$ are square orthogonal matrices ($\vect{U}^{-1} = \vect{U}^T$ and $\vect{V}^{-1} = \vect{V}^T$),
and $\vect{\Sigma}$ is a diagonal matrix with the same dimensions as $\vect{D}$ and with non-negative diagonal entries in decreasing order.
To tell if $\partial f/\partial\vect{p}$ is in the column space of $\vect{D}$, we form $\vect{U}^{T} \, \partial f/\partial\vect{p}$.
The vector elements of $\vect{U}^{T} \, \partial f/\partial\vect{p}$ corresponding to vanishing singular values and to rows $>M$ all are 0
if and only if $\vect{q}$ is in the column space of $\vect{D}$. If so, then equation (\ref{eq:main_result}) can 
be solved by applying the pseudo-inverse of $\vect{D}$:
\begin{equation}
\vect{S} = \vect{V} \vect{\Sigma}^+ \vect{U}^T  \frac{\partial f}{\partial\vect{p}}
\end{equation}
where
$\vect{\Sigma}^+$ is a diagonal rectangular matrix of the same size as $\vect{\Sigma}^T$, with diagonal entries given by
the reciprocal of the corresponding diagonal entries of $\vect{\Sigma}$, except that the reciprocal of 0 entries is replaced by 0. 
The $QR$ and SVD approaches, if all singular values are retained in the latter, give identical results within roundoff error.
The $QR$ decomposition is faster than the SVD, but the singular value spectrum provided by the latter can provide some additional insight.

In a real calculation, $m$ and $n$ cannot extend to infinity,
and some level of discretization error will be present in $\partial f/\partial \vect{p}$ associated with whichever codes are called
to compute it. Therefore, $\partial f/\partial\vect{p}$ will not lie \emph{exactly} in the column space of $\vect{D}$,
but it will be close:
the elements in rows $>M$ of $\vect{Q}^T \, \partial f/\partial\vect{p}$ (in the $QR$ approach) and $\vect{U}^T \, \partial f/\partial\vect{p}$ (in the SVD approach) 
will be much smaller in magnitude than the elements in rows $\le M$.
For the typical case in which $\vect{D}$ has approximately twice as many rows as columns, then the second
half of the elements of $\vect{Q}^T \, \partial f/\partial\vect{p}$ and $\vect{U}^T \, \partial f/\partial\vect{p}$ should be small compared to the first
half.

\begin{figure}[h!]
% Also see 20170917-01 Lab book- figure of normal displacement sensitivity for area.docx
\includegraphics[width=3.5in]{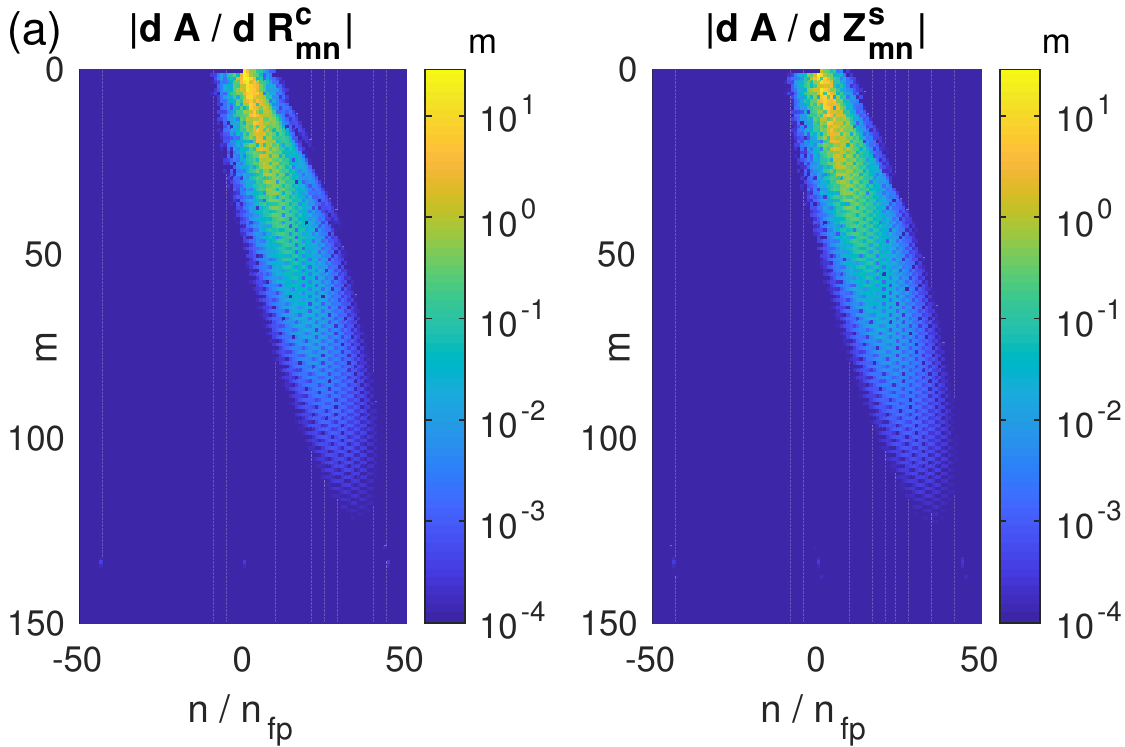}
\includegraphics[width=3.5in]{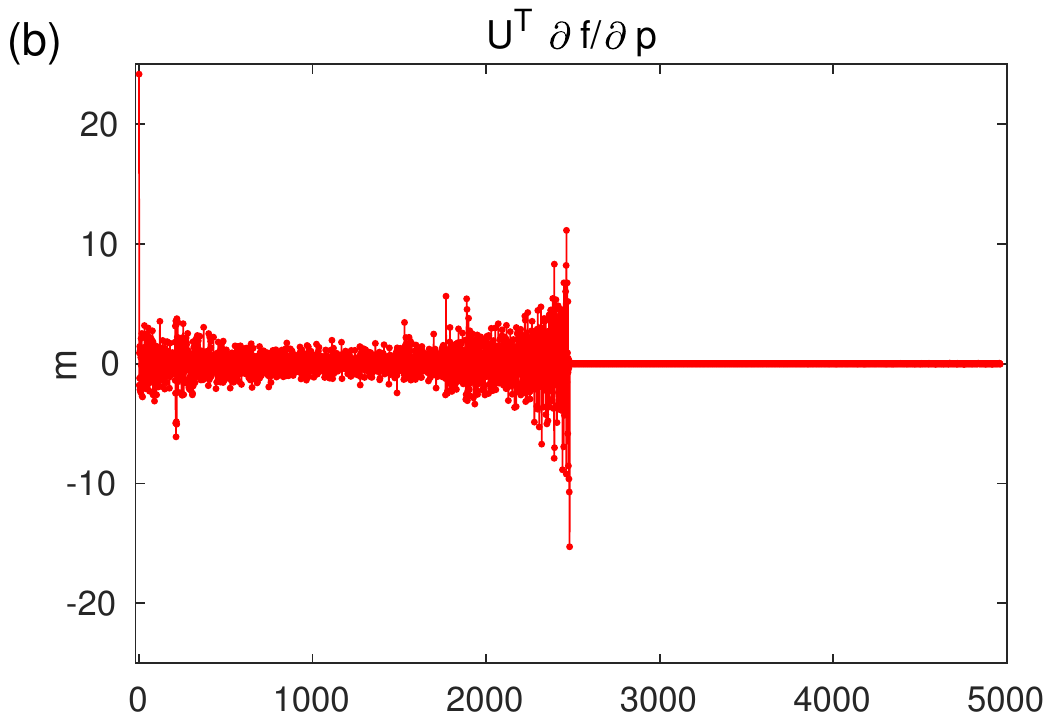}
\\
\vspace{0.2in}
\includegraphics[width=3.5in]{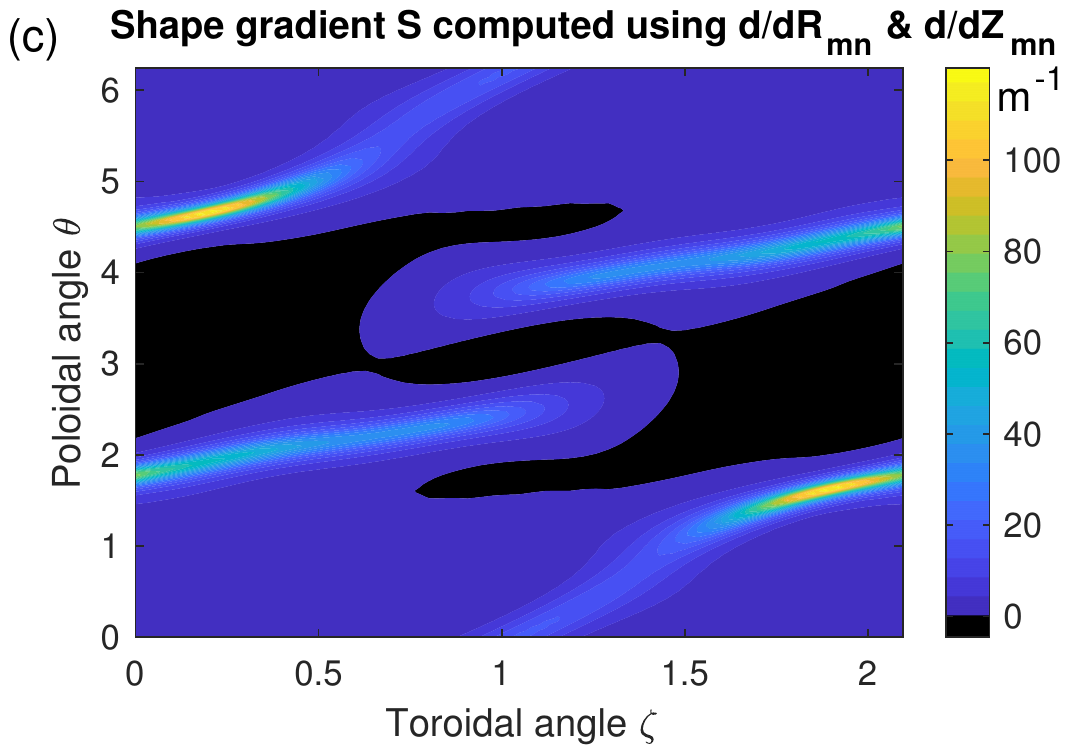}
\caption{(Color online)
Verification of the procedure for computing the shape gradient from parameter derivatives,
using $f=$ surface area in NCSX.
(a) Derivatives of the area with respect to the Fourier harmonics 
of the plasma boundary shape, $\partial f/\partial \vect{p}$, computed using finite differences.
(b) Projection of $\partial f/\partial \vect{p}$ onto the left singular vectors of $\vect{D}$.
The second half of the vector elements are negligible,
confirming that changes to the surface area depend only on the normal component of displacement,
not on the tangential component.
(c) The shape gradient for area computed using the method of section (\ref{sec:finite_difference}).
The result is nearly indistinguishable from figure \ref{fig:sensitivity_area_3D}.a. 
\label{fig:sensitivity_area_Fourier}}
\end{figure}

The procedure described above is illustrated in figure \ref{fig:sensitivity_area_Fourier}. Here, we consider $f=A=$ area, considering NCSX geometry as in section \ref{sec:area} (with $Q=1$),
and the parameter vector $p_j$ is taken to consist of the $R_{m,n}^c$ and $Z_{m,n}^s$ coefficients.
The derivatives $\partial f/\partial \vect{p}$ are computed by finite differences and are shown in figure \ref{fig:sensitivity_area_Fourier}.a.
For this example, we use the same maximum $m$ for $R_{m,n}^c$, $Z_{m,n}^s$, and $S_q$, and the same is true of the maximum $n$, so
$\vect{D}$ has approximately twice as many rows ($\sim 5000$) as columns ($\sim 2500$). 
Unusually large values $m_{\max}=60$ and $n_{\max}=20$ are chosen here to demonstrate numerical stability at this high resolution, and
to resolve the narrow regions of large $H$; indistinguishable results can be obtained with smaller or larger values of $m_{\max}$ and $n_{\max}$.
Figure \ref{fig:sensitivity_area_Fourier}.b shows 
the projection of $\partial f/\partial \vect{p}$
onto the left singular vectors of $\vect{D}$. The second half of these elements are negligible compared to the first half of the elements,
indicating that $\partial f/\partial \vect{p}$ lies in the column space of $\vect{D}$, and so (\ref{eq:main_result}) is solvable.
Therefore a shape gradient $S$ exists.
The ratio of largest to smallest singular values of $\vect{D}$ is only 11.0, so the solution is well conditioned even at this quite high resolution.
The solution for $S$ is then displayed in figure \ref{fig:sensitivity_area_Fourier}.c.
The result is nearly indistinguishable from figure \ref{fig:sensitivity_area_3D}.a, verifying that the procedure produces the correct result.

In \cite{Elizabeth}, figure 9 displays shape gradients computed using the method described here.
That figure shows the error in producing a desired plasma shape is highly sensitive to the plasma-coil distance
near concave regions of the plasma boundary, as is the complexity of the coil shapes needed to produce the desired plasma shape.

The method described here for computing shape gradient for surfaces can also be applied
to coils, as an alternative to the method of section \ref{sec:finite_differences_coils}.
Instead of considering all three components of $\vect{S}$ as unknowns,
we write
$\vect{S} = S_1 \vect{v}_1 + S_2 \vect{v}_2$ where $\vect{v}_1$ and $\vect{v}_2$ are two vectors defined at each point
along the coil that span the plane normal to the curve. For instance, one could take $\vect{v}_1$ and $\vect{v}_2$ to be
the Frenet-Serret normal and binormal vectors, or one could take $\vect{v}_1$ to be $\vect{e}_\zeta \times \vect{t}$, the cylindrical azimuthal
direction crossed with the tangent vector, and $\vect{v}_2 = \vect{t}\times\vect{v}_1$.
Expanding  $S_1$ and $S_2$ in Fourier series analogous to (\ref{eq:S_coil_Fourier}), 
the vector of Fourier amplitudes 
$\vect{S}^{coil} = (S_{1,m}^c, S_{1,m}^s, S_{2,m}^c, S_{2,m}^s)^T$
becomes the unknown we seek. The discretized system (\ref{eq:Sdef3Dcurves_alt})
then has $3/2$ as many equations as unknowns. The remaining steps of the calculation are the same
as for the surface case. For all the examples of coil shape gradients in this paper, it was verified that 
this alternative procedure gave identical results to the procedure of section \ref{sec:finite_differences_coils},
within a small discretization error.

%%%%%%%%%%%%%%%%%%%%%%%%%%%%%%%%%%%%%%%%%%%%%%%%%%%%%%
%%%%%%%%%%%%%%%%%%%%%%%%%%%%%%%%%%%%%%%%%%%%%%%%%%%%%%

\section{Coil tolerances}
\label{sec:tolerance}

Once the shape gradient for coils has been computed, practical tolerances on the coil shapes can be derived as follows.
First, an acceptable variation in the figure of merit is identified, $\Delta f$, which is positive. Then a local tolerance $T_k(\ell)$ for each coil $k$ can be computed from
\begin{equation}
T_k(\ell) = \frac{w_k(\ell) \, \Delta f}{\sum_{k'} \int d\ell' \, w_{k'}(\ell') |\vect{S}_{k'}(\ell')|},
\label{eq:tolerance}
\end{equation}
where $\ell$ denotes distance along the coil, and $w_{k}(\ell)$ is a weighting function that is arbitrary, except it must be non-negative everywhere and it must have a positive integral on at least one coil. 
We can then prove that if the magnitude of the shape perturbation satisfies $|\delta \vect{r}| \le T_k$
at all points along the coils, the perturbation magnitude $|\delta f|$ will be $\le \Delta f$ (in the approximation that variations are linear):
\begin{equation}
|\delta f| \le \sum_k \int d\ell |\vect{S}_k\cdot\delta\vect{r} |
\le  \sum_k \int d\ell |\vect{S}_k| |\delta\vect{r} |
\le \sum_k \int d\ell |\vect{S}_k| T_k
= \Delta f.
\end{equation}
The weight $w_k$ allows the tolerance to be distributed in different ways among and along the coils. 
The choice $w_k = |\vect{S}_k|$ makes the tolerance spatially uniform, $T_k = \Delta f / \left( \sum_{k'} \int d\ell' |\vect{S}_{k'}(\ell')| \right)$.
If a design includes both non-planar and planar coils, it may make sense to choose $w_k$ larger on the former than on the latter,
allocating a larger fraction of the total
$\Delta f$ to the more difficult-to-built non-planar coils.
The choice $w_k = |\vect{S}_k|^{-\alpha}$ for some $\alpha \ge 0$ can be useful, as it
results in a tighter tolerance in the coil regions to which the plasma is most sensitive.
For any weight, the tolerance (\ref{eq:tolerance}) is conservative
in that it provides a limit on the worst possible $\delta f$, rather than a limit on the 
$\delta f$ expected based on some distribution of likely errors. 
The uniform tolerance obtained from $w_k=1$ could potentially be included in an optimization
objective function, with the goal of obtaining robust stellarator designs that have large tolerances.

%%%%%%%%%%%%%%%%%%%%%%%%%%%%%%%%%%%%%%%%%%%%%%%%%%%%%%
%%%%%%%%%%%%%%%%%%%%%%%%%%%%%%%%%%%%%%%%%%%%%%%%%%%%%%
%%%%%%%%%%%%%%%%%%%%%%%%%%%%%%%%%%%%%%%%%%%%%%%%%%%%%%
%%%%%%%%%%%%%%%%%%%%%%%%%%%%%%%%%%%%%%%%%%%%%%%%%%%%%%

\section{Example: Rotational transform}
\label{sec:iota}

\begin{figure}[h!]
\includegraphics[width=3.5in]{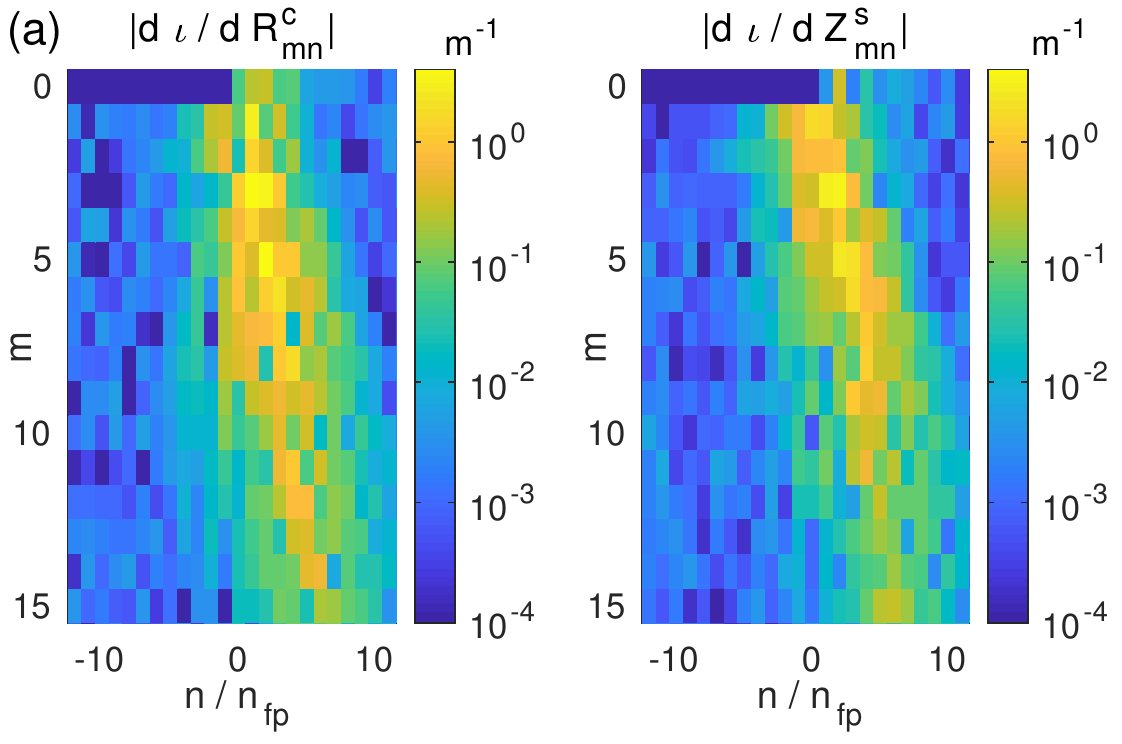}
\includegraphics[width=3.5in]{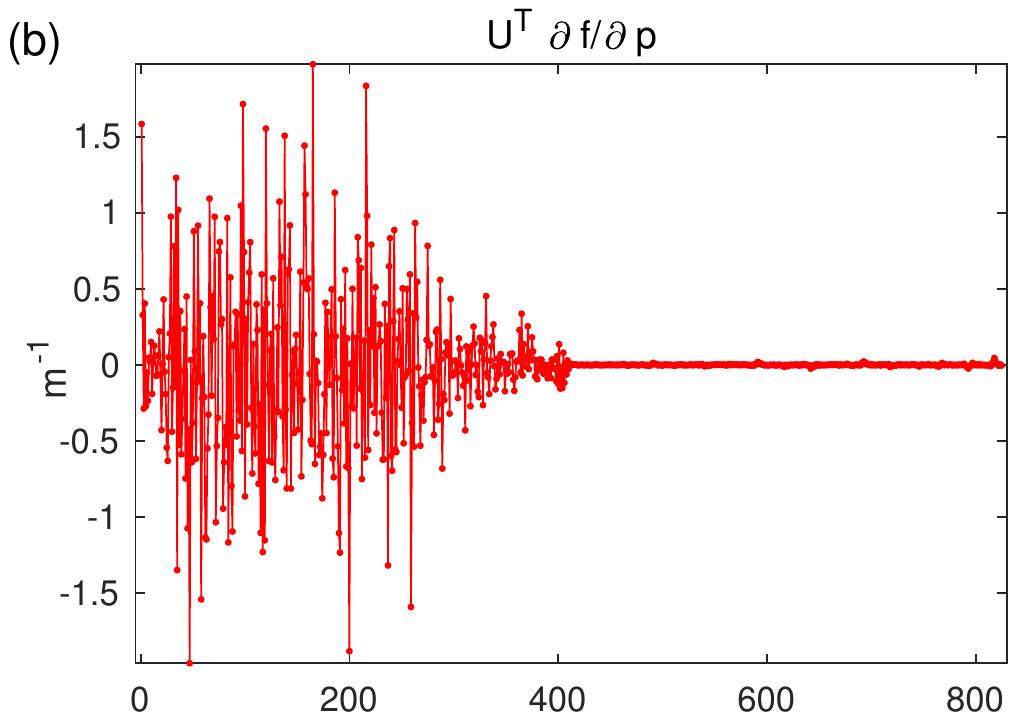}
\\
\vspace{0.2in}
\includegraphics[width=3.5in]{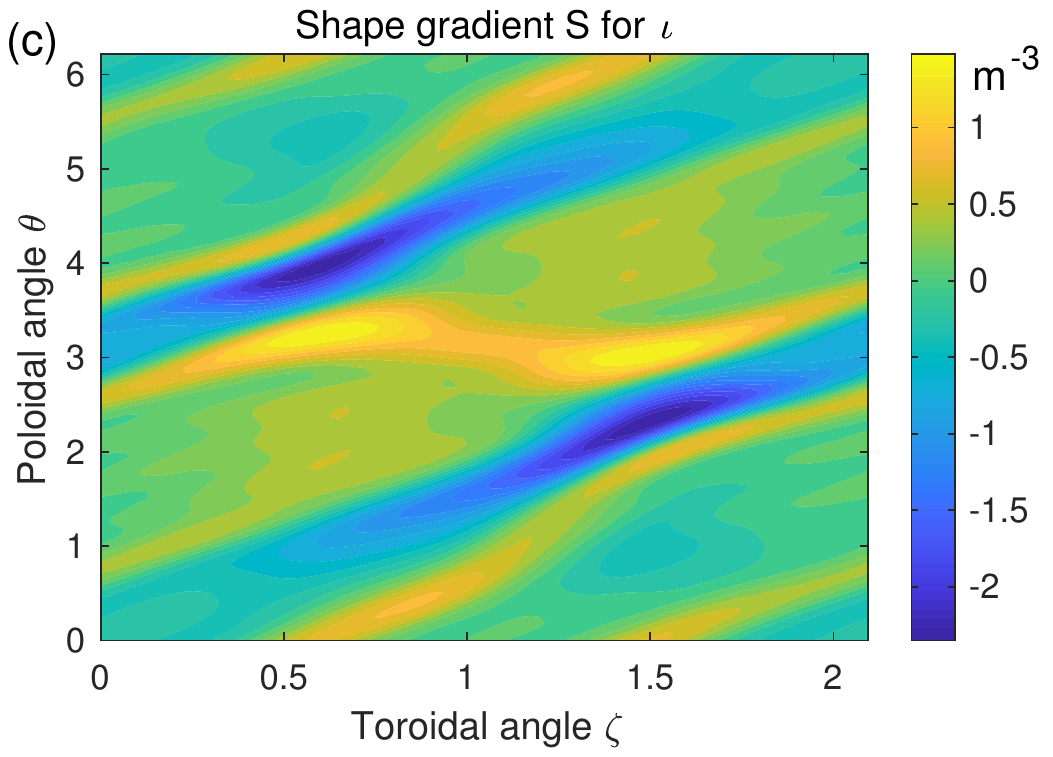}
\includegraphics[width=3.5in]{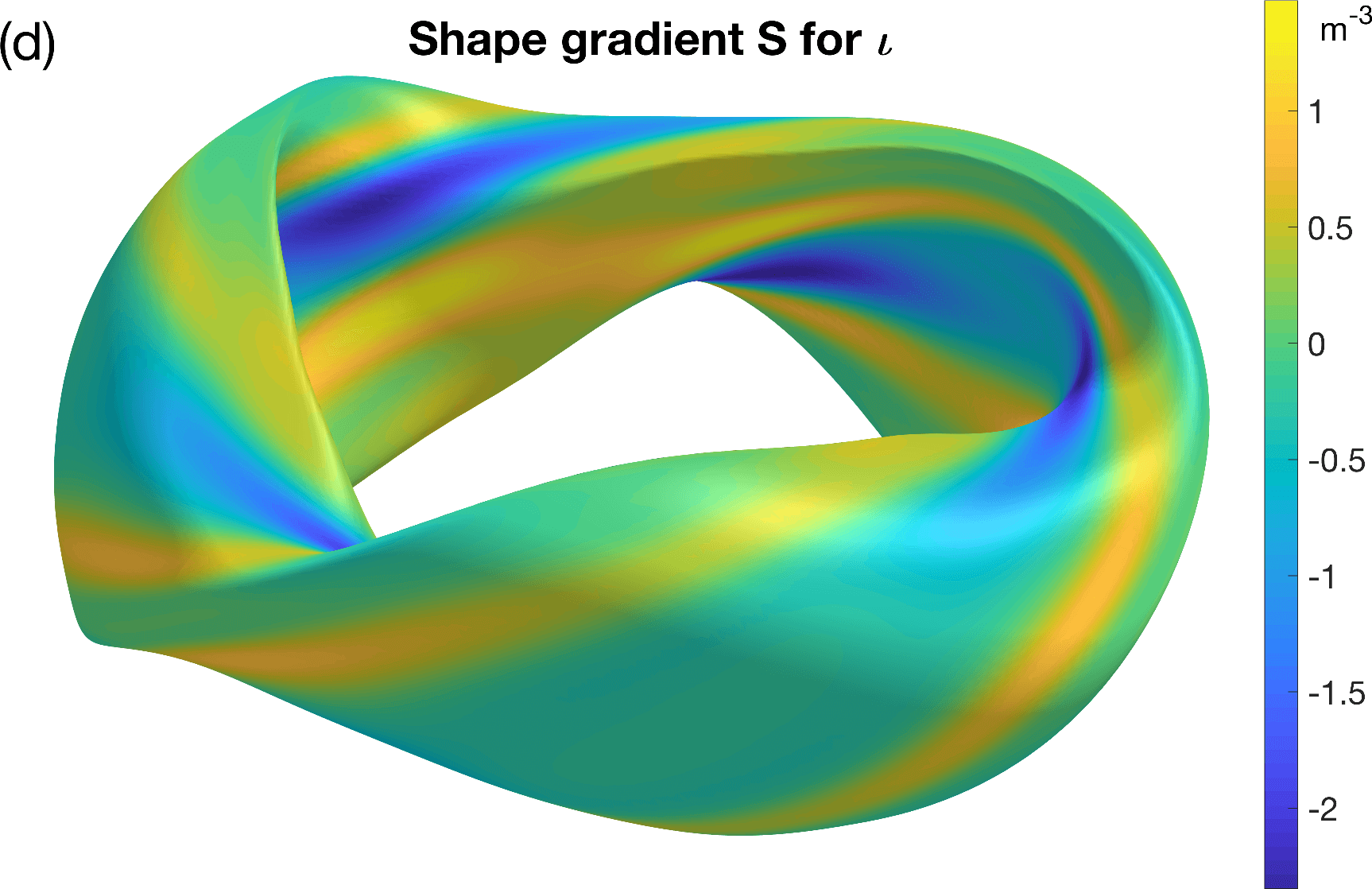}
\caption{(Color online)
Computation of the shape gradient for the half-radius rotational transform $\iota$.
(a) Derivatives of $\iota$ with respect to the Fourier harmonics 
of the plasma boundary shape, computed with finite differences using STELLOPT.
(b) Elements of the vector $\vect{U}^{T} \, \partial f/\partial\vect{p}$. The second half of the vector elements are negligible
compared to the first half,
confirming that changes to $\iota$ depend only on the normal component of displacement,
not on the tangential component.
(c) The shape gradient for $\iota$ computed using the method of section (\ref{sec:finite_difference}).
(d) The same quantity viewed in three dimensions.
\label{fig:sensitivity_iota_Fourier}}
\end{figure}

Now let us apply  the procedures of the previous section to calculate shape gradient for a figure of merit for which
an analytic expression for the shape gradient is unavailable:
the rotational transform $\iota$ in NCSX at half radius, $\sqrt{\psi/\psi_a}=0.5$ where $2\pi\psi$ is the toroidal flux
and $\psi_a$ is the value of $\psi$ at the plasma boundary. 
We will calculate the  shape gradient for both the plasma boundary shape and coil shapes.
Beginning with the former, derivatives of $\iota$ with respect to the Fourier modes
of the plasma boundary, $d \iota / d R_{m,n}^c$ and $d \iota / d Z_{m,n}^s$, are computed using finite differences. To compute these finite differences, the STELLOPT code \cite{stellopt_Spong,stellopt_Reiman}
is used to run the VMEC code \cite{VMEC1983, VMEC1986} (in fixed-boundary mode) many times for slightly different plasma boundary shapes. 
(STELLOPT was modified slightly to evaluate centered rather than one-sided differences.)
The resulting derivatives are displayed
in figure \ref{fig:sensitivity_iota_Fourier}.a.
We adopt the VMEC conventions that $m \ge 0$, and for $m=0$, $R_{mn}^c$ is nonzero only for $n \ge 0$ and $Z_{mn}^s$ is nonzero only for $n>0$.
We verified that derivatives of $\iota$ with respect to $R_{m,n}^s$ and $Z_{m,n}^c$ vanish for all $m$ and $n$, as do derivatives with respect to
$R_{m,n}^c$ and $Z_{m,n}^s$ when $n$ is not a multiple of 3, so $S$ has stellarator symmetry and $n_{fp}$ symmetry by the argument in appendix
\ref{sec:symmetry}.

Using these derivatives, we form the vector $\partial f/\partial\vect{p}$, and project it onto the left singular vectors of $\vect{D}$, yielding figure \ref{fig:sensitivity_iota_Fourier}.b.
The second half of the resulting vector elements have negligible magnitude, confirming that perturbations to $\iota$ depend only on the normal displacement
and not on tangential displacements, and so a shape gradient can be constructed.
Completing the procedure of section \ref{sec:finite_differences_surfaces}, we obtain the shape gradient  $S$ in figures \ref{fig:sensitivity_iota_Fourier}.c-d.
In these figures it can be seen that the shape gradient is small in magnitude in the bean cross section ($\zeta=0$)
and the bullet cross section ($\zeta = \pi/3$), and is maximum in magnitude in between these cross sections on the inboard side.

For the calculations shown here, the nonzero plasma current associated with the
$\beta=4\%$ NCSX target design has been retained.
For simplicity, in this paper we will take the radial profiles of pressure and toroidal current (expressed as functions of toroidal flux normalized
to the boundary toroidal flux) to be fixed as the boundary is deformed. (More sophisticated options are available:
the two profile functions could be understood to depend implicitly on the boundary shape via equations for radial transport and the bootstrap current,
or terms representing contributions to $\delta f$ from independent perturbations to the profile functions could be added to (\ref{eq:Sdef3D}).)
We have also computed the shape gradient for $\iota$ using the same boundary shape at zero plasma pressure and zero plasma current,
and it is nearly indistinguishable from figures \ref{fig:sensitivity_iota_Fourier}.c-d. We have also verified that all results in figure \ref{fig:sensitivity_iota_Fourier} are converged
with respect to the VMEC resolution parameters (number of radial grid points {\ttfamily ns}, number of poloidal modes {\ttfamily mpol},
number of toroidal modes {\ttfamily ntor}), number of grid points in $\theta$ and $\zeta$ used for the integrations in section \ref{sec:finite_differences_surfaces}, and the number of poloidal and toroidal modes used for the procedure of 
section \ref{sec:finite_differences_surfaces}.

\begin{figure}[h!]
\includegraphics[width=3.5in]{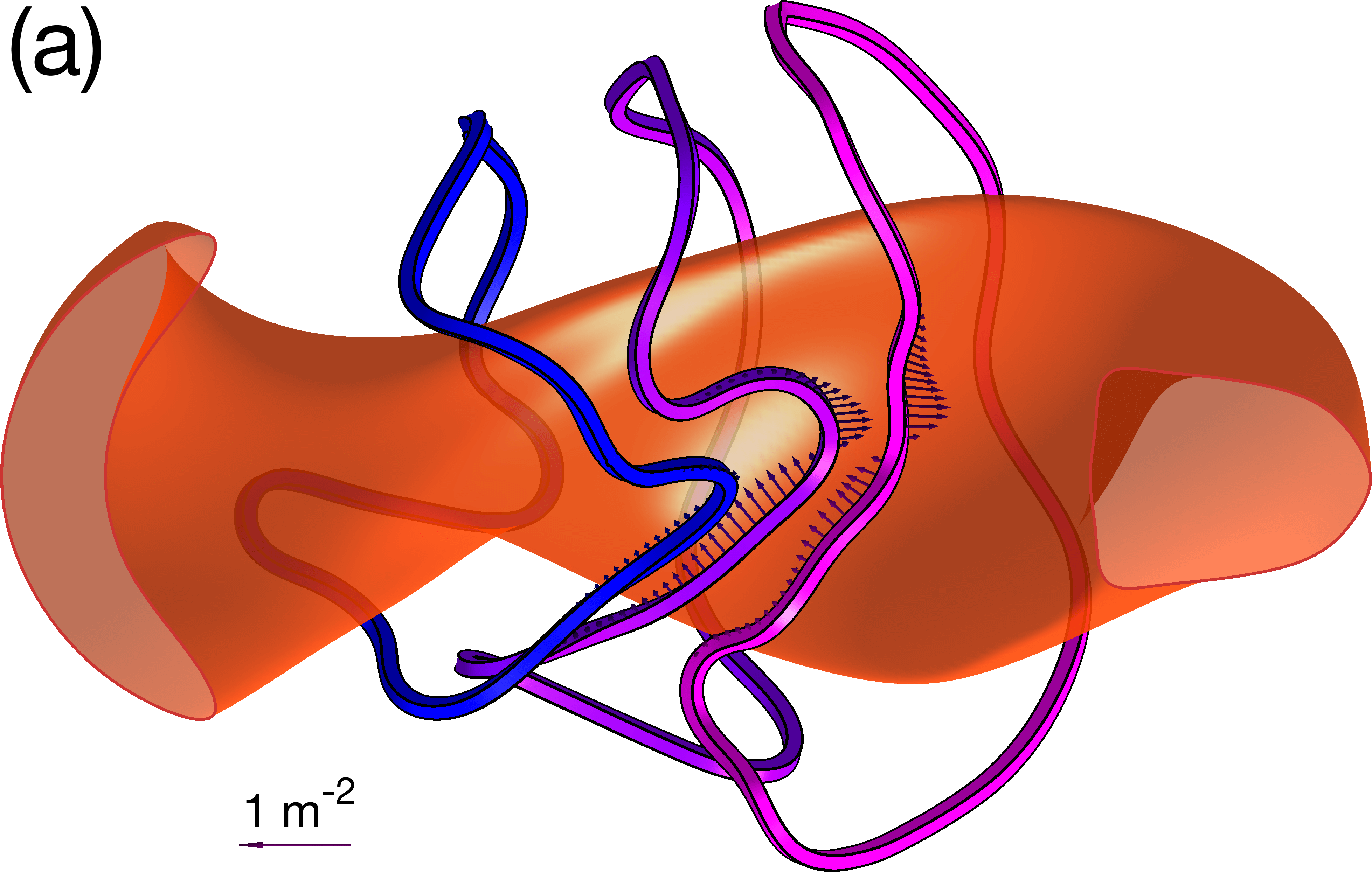}
\includegraphics[width=3.5in]{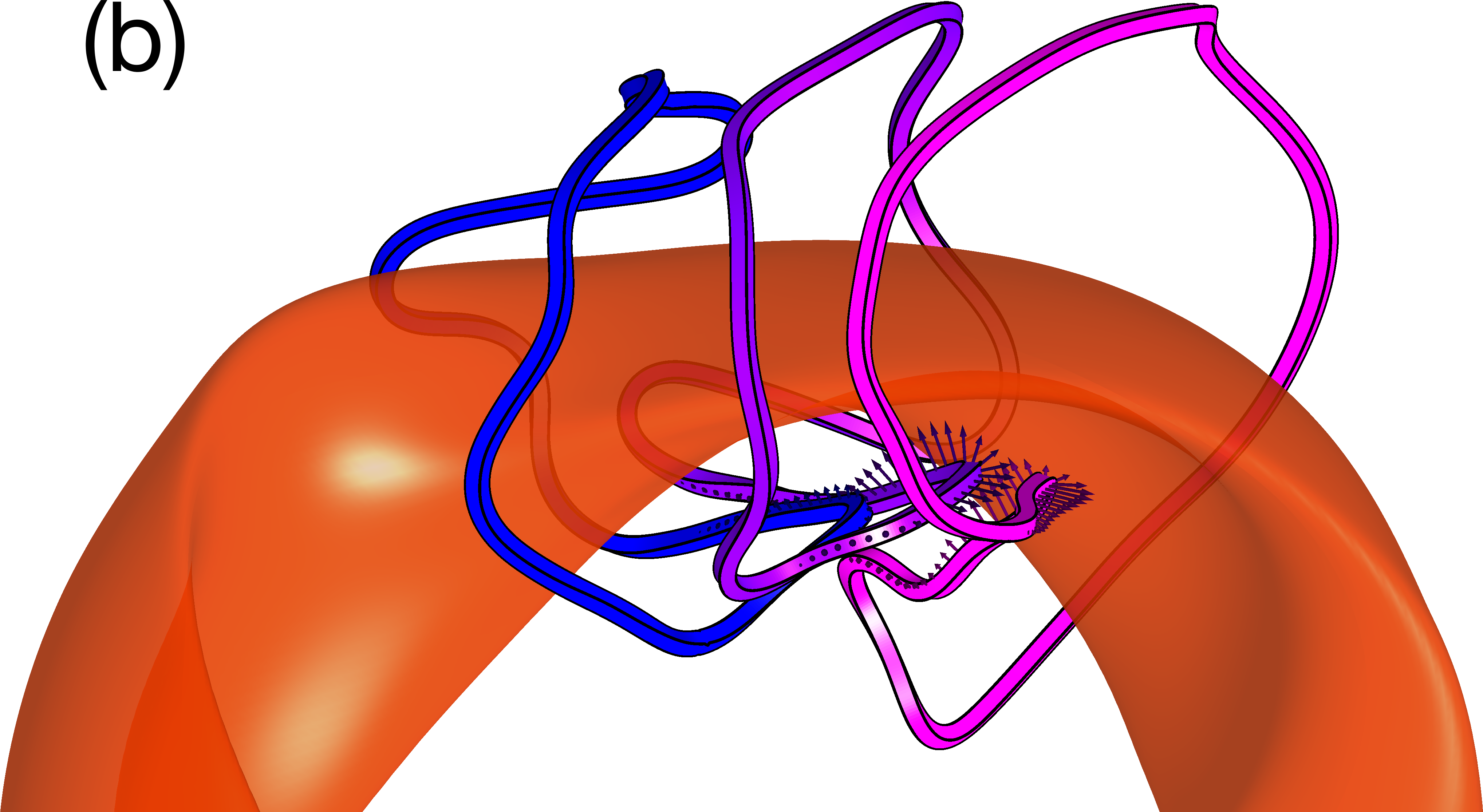}
\includegraphics[width=3.5in]{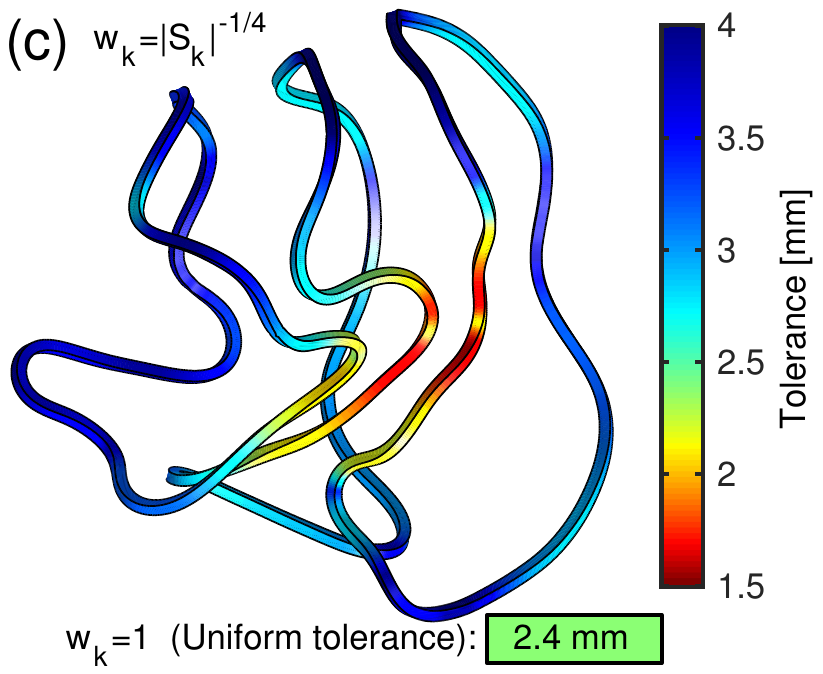}
\caption{(Color online)
Arrows show the shape gradient for $\iota$ at half-radius for the NCSX coils. Panels (a) and (b) show the same data from two different angles.
The plasma boundary is shown in red. Only one of the six sets of three unique coil shapes is shown.
The region of significant gradient is highly localized to the inboard side.
(c) The shape tolerance required to achieve $\Delta \iota \le 0.02$ computed from the shape gradient
using (\ref{eq:tolerance}), using two different choices for the weight $w_k$.
\label{fig:sensitivity_iota_coils}}
\end{figure}

Next, in figure \ref{fig:sensitivity_iota_coils} we show the shape gradient $\vect{S}$ for \emph{coils} for the same quantity $\iota$.
To perform this calculation, we use STELLOPT to call VMEC many times in free-boundary mode. STELLOPT varies
the Fourier amplitudes (\ref{eq:coil_Fourier}) of the Cartesian components of the coil shapes, and in each case the magnetic field from the perturbed coil
shapes is calculated using the MAKEGRID code, which is then provided as an input to free-boundary VMEC. The resulting finite difference derivatives
are then used as the right-hand side for the algorithm of section \ref{sec:finite_differences_coils}.
Since the surface shape gradient in figure \ref{fig:sensitivity_iota_Fourier}.c-d is stellarator-symmetric and $n_{fp}$-symmetric, we expect the coil
shape gradient to have these same symmetries, and we have verified these symmetries numerically.
Specifically, $\delta \iota$ when a coil is perturbed in unison with its stellarator-symmetric partner is exactly twice $\delta \iota$ when 
only one of the pair is perturbed, and $\delta \iota$ when a coil is perturbed in unison with its two $n_{fp}$-symmetric partners is exactly 
three times $\delta \iota$ when only one of the triplet is perturbed. These symmetries are quite useful as they mean that VMEC can be run
assuming stellarator symmetry and $n_{fp}$ symmetry, reducing the computational expense.

The shape gradient, coils, and plasma boundary are shown in figure \ref{fig:sensitivity_iota_coils} from two different perspectives: 
from the $Z$ axis looking outward, and a bird's eye view.
The shape gradient $\vect{S}_k$ is displayed using arrows.
Note that the scale for the arrow lengths is independent of the scale for the other objects.
Over much of the coils, the shape gradient is small enough that these arrows do not extend outside the coils, the width of which is chosen
for convenient visualization and differs from the actual coil thickness.
It was verified that the shape gradient is orthogonal to the coils, as it should be.
It is apparent that the shape gradient is highly localized to the inboard side of the coils. 
The shape gradient is also larger for the type A and B coils (magenta and purple respectively) that are closer to the bean-shaped plasma cross-section than for the type C coils (blue) that are closer to the bullet-shaped cross section. This localization is likely due in part to the fact that the coils are closer
to the plasma in the regions of large shape gradient, so displacements of the coils in these regions cause a relatively large change to the magnetic field in the plasma region.

The coil shape tolerance (\ref{eq:tolerance}) is displayed in figure \ref{fig:sensitivity_iota_coils}.c. For this computation,
we arbitrarily choose the acceptable variation in the figure of merit to be $\Delta \iota=0.02$.
We also choose the weight $w_k$ to be zero on all the planar coils, effectively focusing only on the contribution
to $\Delta \iota$ from the modular coils. 
The bottom of figure \ref{fig:sensitivity_iota_coils}.c shows the uniform tolerance
calculated using (\ref{eq:tolerance}) with $w_k = 1$.
For comparison, the main part of figure \ref{fig:sensitivity_iota_coils}.c shows a nonuniform tolerance
that gives the same bound on $\delta \iota$, calculated using $w_k = |\vect{S}_k|^{-1/4}$.
By making the weight $w_k$ scale inversely with $|\vect{S}_k|$ in this way, the tolerance
can be relaxed over a majority of the coils if the tolerance is tightened in a few small regions.
The exponent $-1/4$ was chosen since it gives a reasonable range for the tolerance; any other value is
equally valid mathematically.
This nonuniform tolerance is tightest where the shape gradient is largest in magnitude, on the inboard
side, particularly near the bean cross section. Figure \ref{fig:sensitivity_iota_coils}.c makes it  apparent that
it is most important to rigidly support the coils on the inboard side to minimize variation in this figure of merit.

%%%%%%%%%%%%%%%%%%%%%%%%%%%%%%%%%%%%%%%%%%%%%%%%%%%%%%
%%%%%%%%%%%%%%%%%%%%%%%%%%%%%%%%%%%%%%%%%%%%%%%%%%%%%%
%%%%%%%%%%%%%%%%%%%%%%%%%%%%%%%%%%%%%%%%%%%%%%%%%%%%%%
%%%%%%%%%%%%%%%%%%%%%%%%%%%%%%%%%%%%%%%%%%%%%%%%%%%%%%

\section{Example: Neoclassical transport}

\begin{figure}[h!]
\includegraphics[width=3.5in]{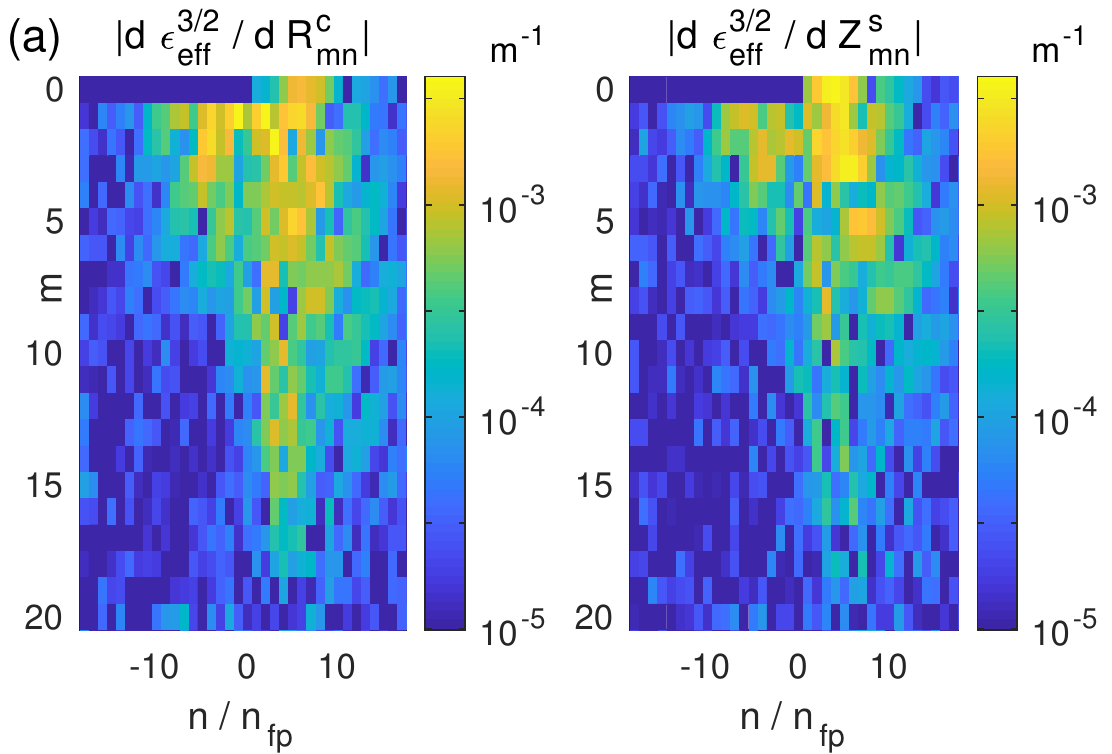}
\includegraphics[width=3.5in]{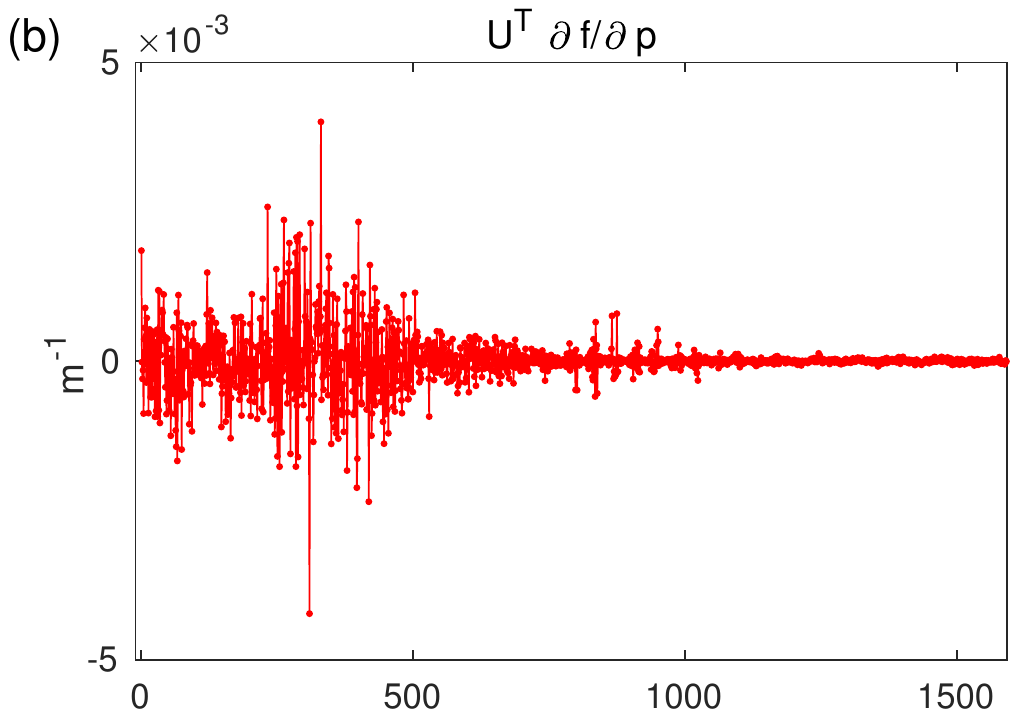}
\\
\vspace{0.2in}
\includegraphics[width=3.5in]{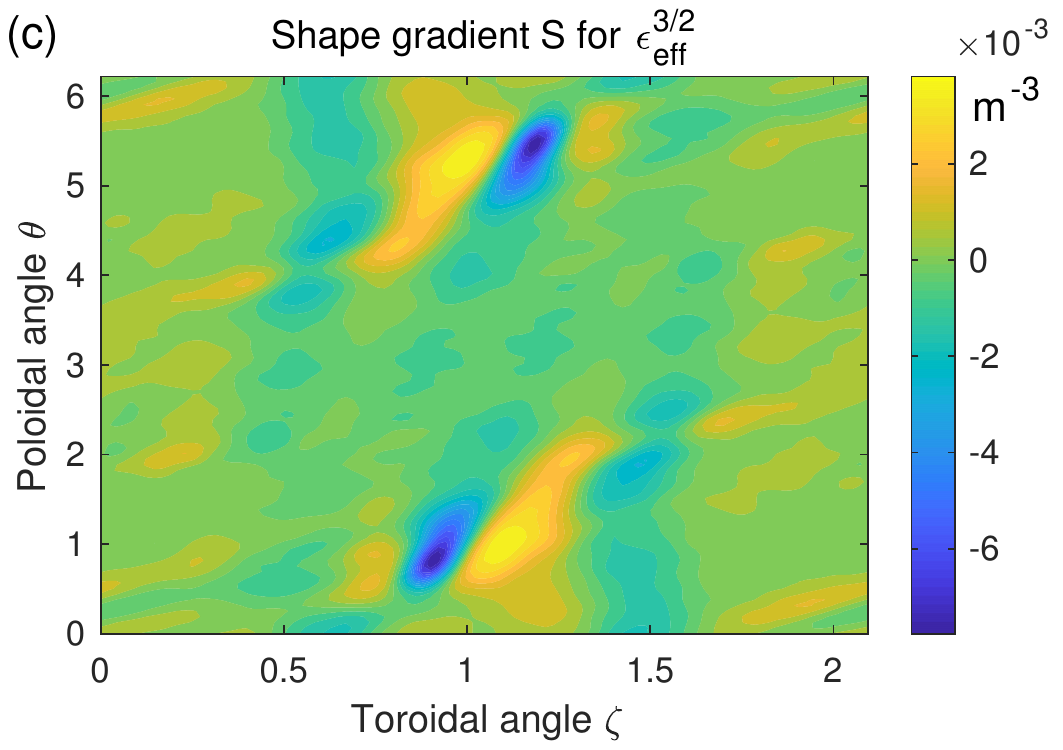}
\includegraphics[width=3.5in]{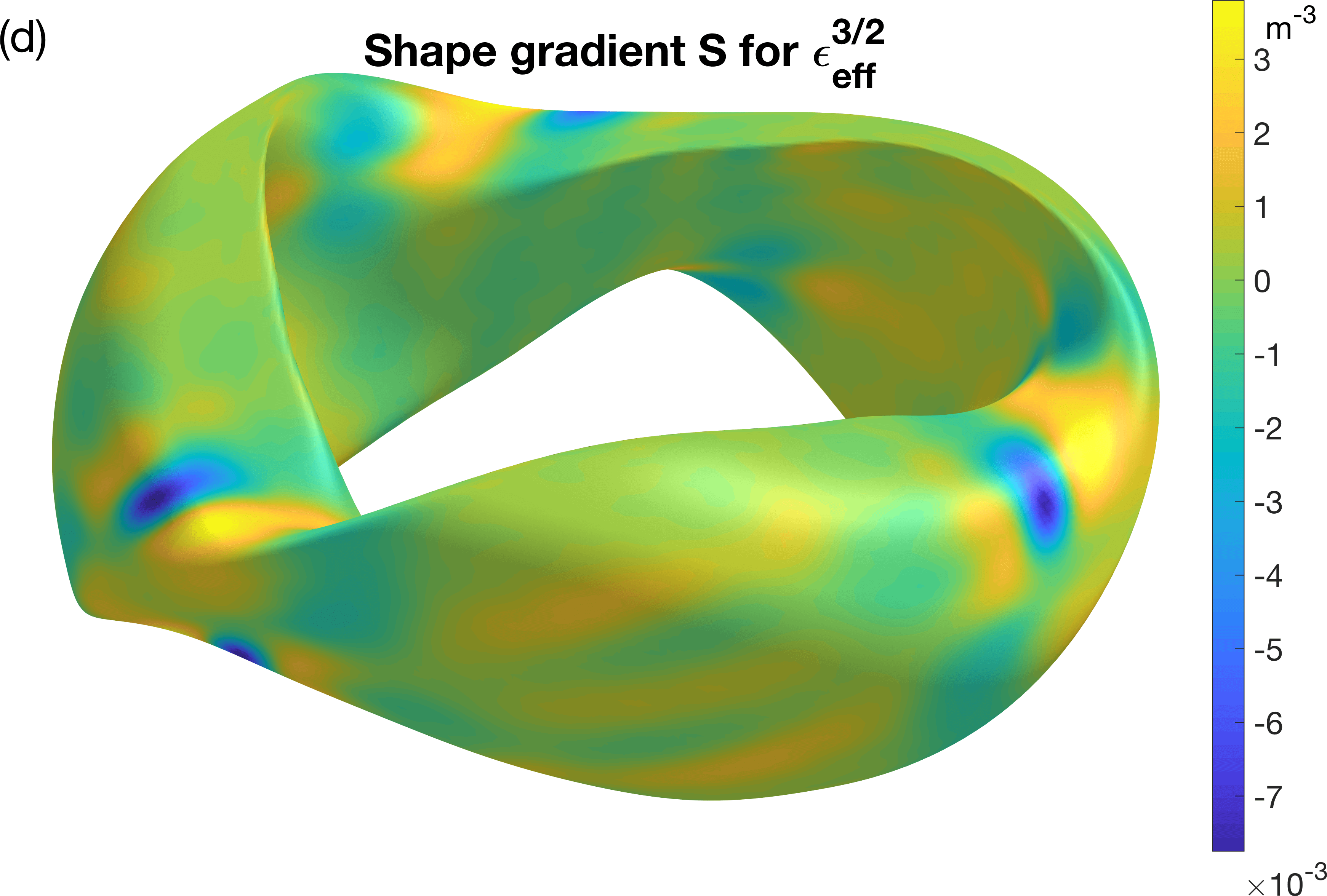}
\caption{(Color online)
Computation of the shape gradient for the half-radius neoclassical transport $\epsilon_{eff}^{3/2}$.
(a) Derivatives of $\epsilon_{eff}^{3/2}$ with respect to the Fourier harmonics 
of the plasma boundary shape, computed with finite differences using STELLOPT.
(b) Projection of $\partial \epsilon_{eff}^{3/2}/\partial\vect{p}$ onto the left singular vectors of $\vect{D}$.
The second half of the elements are small compared to the first,
confirming that changes to $\epsilon_{eff}^{3/2}$ depend only on the normal component of displacement,
not on the tangential component.
(c) The shape gradient for $\epsilon_{eff}^{3/2}$ computed using the method of section (\ref{sec:finite_difference}).
(d) The same quantity viewed in three dimensions.
\label{fig:sensitivity_NEO_Fourier}}
\end{figure}

As a second example, we consider $f=$ the normalized neoclassical transport figure of merit $\epsilon_{eff}^{3/2}$,
again evaluated at half radius in NCSX. 
Derivatives of $\epsilon_{eff}^{3/2}$ with respect to the Fourier amplitudes of the plasma boundary shape are evaluated using the STELLOPT code. STELLOPT calls VMEC for many slightly different boundary shapes, and in each case, STELLOPT then calls the BOOZ\_XFORM code to convert the resulting magnetic equilibrium to Boozer coordinates, and then calls the NEO code \cite{Nemov} to compute $\epsilon_{eff}^{3/2}$.
These derivatives are displayed in figure  \ref{fig:sensitivity_NEO_Fourier}.a.
As with the earlier example, we verified derivatives with respect to $R_{m,n}^s$ and $Z_{m,n}^c$ vanish for all $m$ and $n$, as do derivatives with respect to
$R_{m,n}^c$ and $Z_{m,n}^s$ when $n$ is not a multiple of 3, so $S$ has stellarator symmetry and $n_{fp}$ symmetry by the argument in appendix
\ref{sec:symmetry}.

The remainder of figure \ref{fig:sensitivity_NEO_Fourier} shows the calculation of shape gradient 
with respect to the plasma boundary using the method of section \ref{sec:finite_differences_surfaces}.
In the projection of $\partial f/\partial \vect{p}$ onto the left singular vectors of $\vect{D}$,
shown in figure \ref{fig:sensitivity_NEO_Fourier}.b, the second half of the elements are small compared to the first,
consistent with $\epsilon_{eff}^{3/2}$ being independent of tangential displacements.
There are some small nonzero amplitudes that can be attributed to discretization error.
The final shape gradient, displayed in \ref{fig:sensitivity_NEO_Fourier}.c-d, is evidently most significant at
the top and bottom of the bullet cross section. The reasons for this result would be interesting to explore in future work.
We verified the results in figure  \ref{fig:sensitivity_NEO_Fourier} were converged with respect to the numerical resolution parameters
in the codes involved, and to the number of poloidal and toroidal Fourier modes retained in the shape gradient.

\begin{figure}[h!]
\includegraphics[width=3.5in]{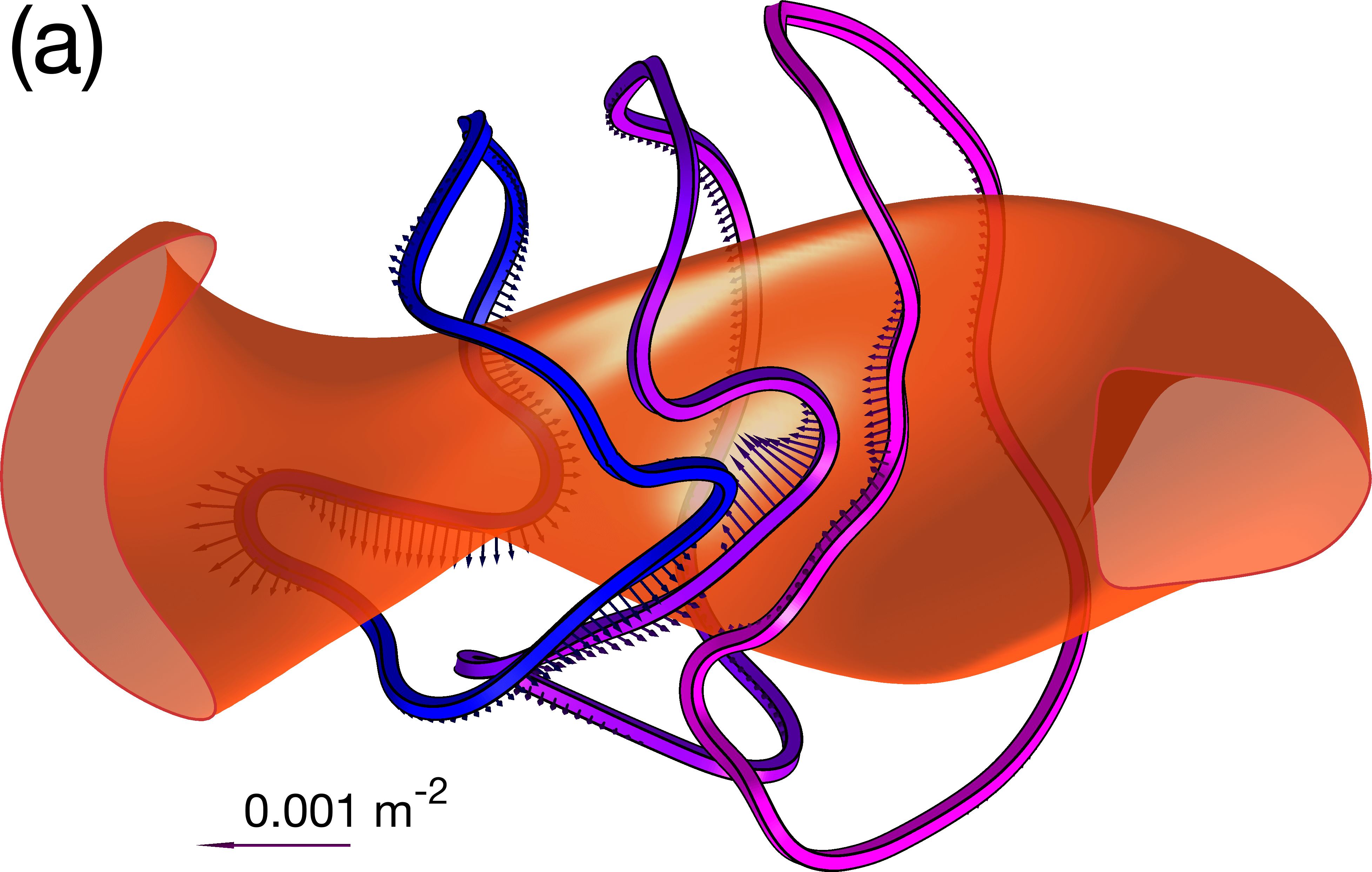}
\includegraphics[width=3.5in]{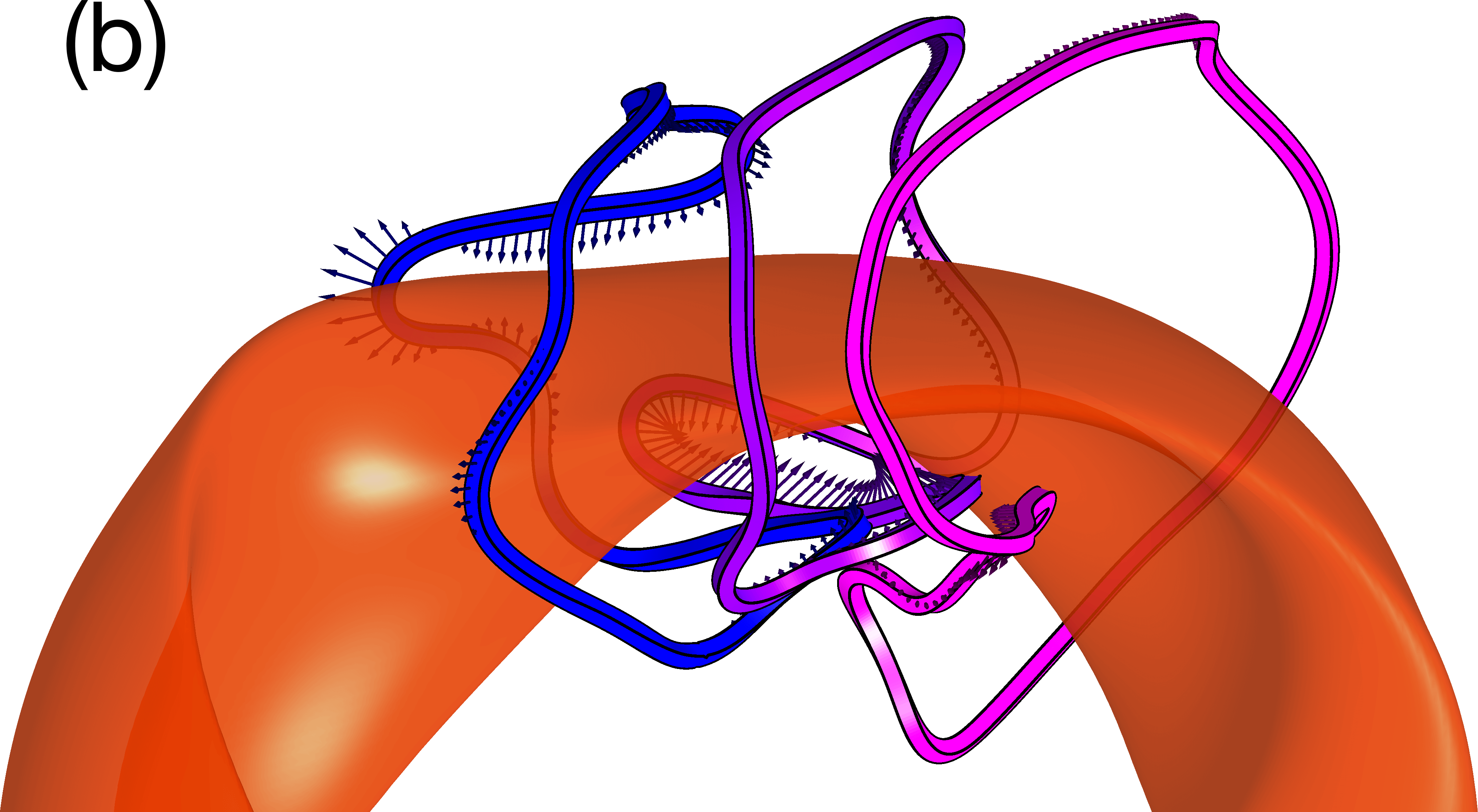}
\\
\includegraphics[width=3.5in]{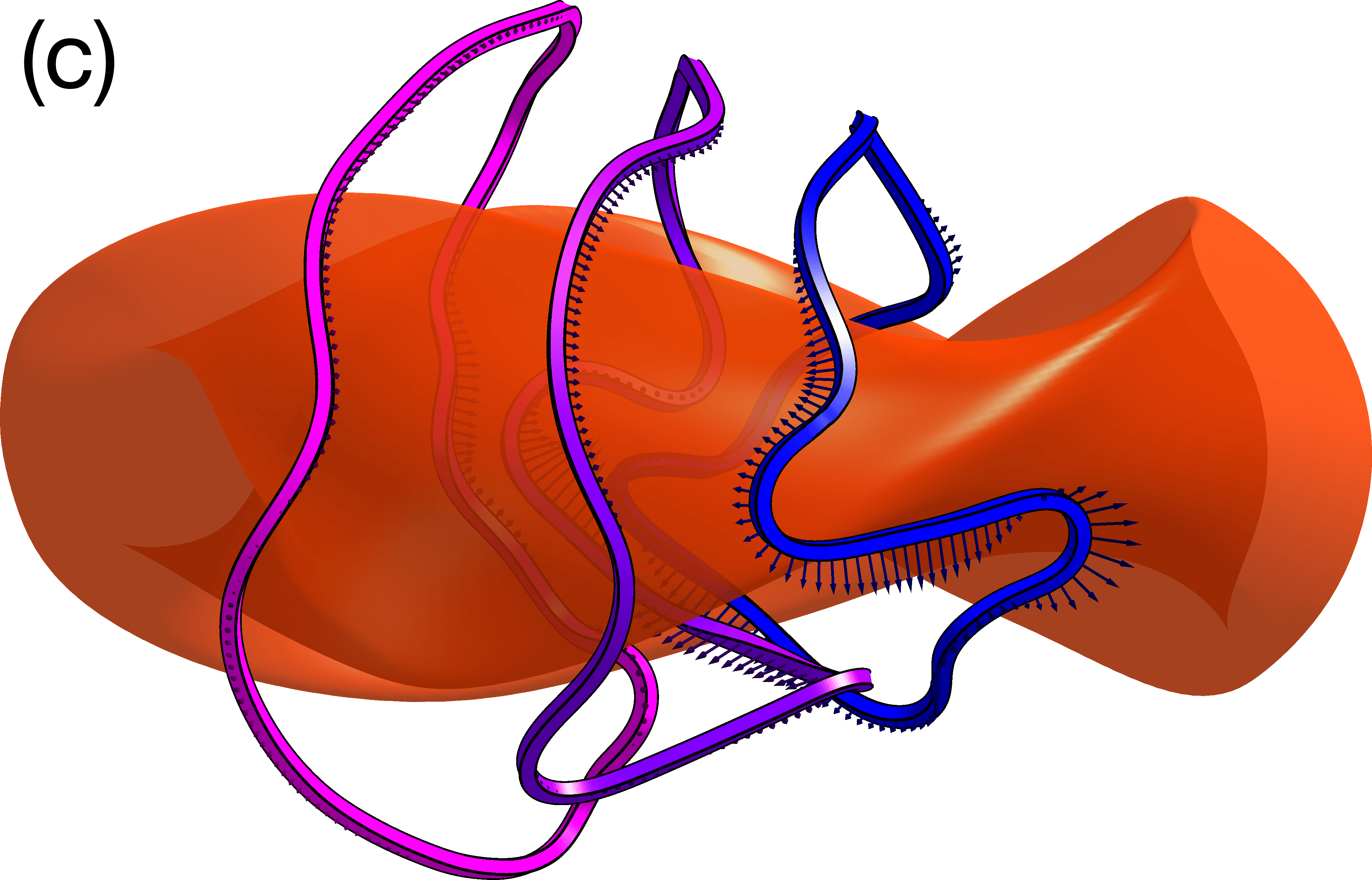}
\includegraphics[width=3.5in]{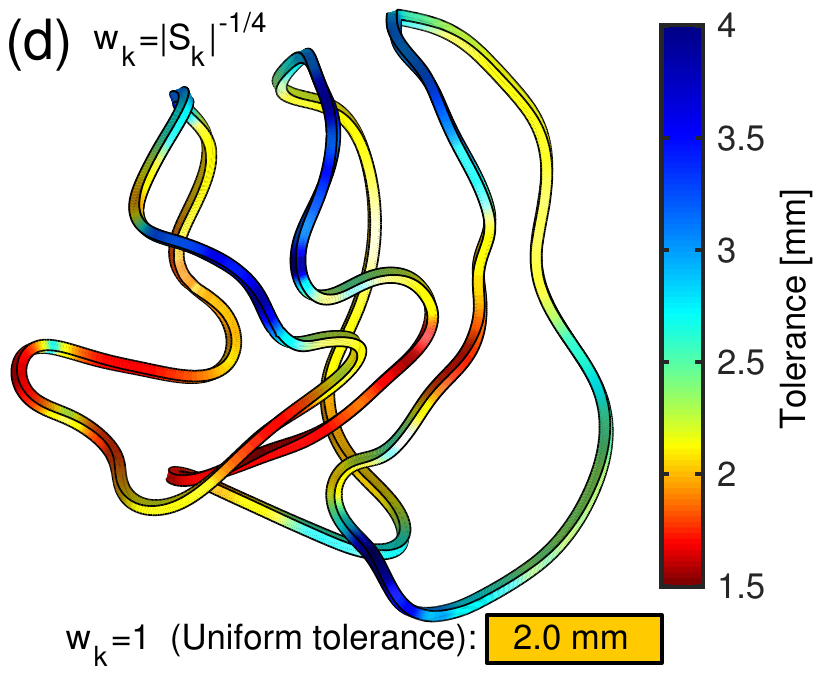}
\caption{(Color online)
Arrows show the shape gradient for $\epsilon_{eff}^{3/2}$ at half-radius for the NCSX coils. Panels (a)-(c) show the same data from different angles.
The plasma boundary is shown in red. Only one of the six sets of three unique coil shapes is shown.
(d) The shape tolerance required to achieve $\Delta \epsilon_{eff}^{3/2} / \epsilon_{eff}^{3/2} \le 1/2$ computed from the shape gradient
using (\ref{eq:tolerance}), for two different choices of the weight $w_k$.
\label{fig:sensitivity_NEO_coils}}
\end{figure}

Next, again considering $f=\epsilon_{eff}^{3/2}$, figure \ref{fig:sensitivity_NEO_coils} shows the shape gradient for coils.
As in section \ref{sec:iota}, the finite difference derivatives are obtained by having STELLOPT call MAKEGRID
and then free-boundary VMEC, this time followed by BOOZ\_XFORM and NEO.
As for the $\iota$ example, we verified numerically that
$\delta \epsilon_{eff}^{3/2}$ when a coil is perturbed in unison with its stellarator-symmetric partner is exactly twice $\delta \epsilon_{eff}^{3/2}$ when 
only one of the pair is perturbed, and $\delta \epsilon_{eff}^{3/2}$ when a coil is perturbed in unison with its two $n_{fp}$-symmetric partners is exactly 
three times $\delta \epsilon_{eff}^{3/2}$ when only one of the triplet is perturbed, consistent with the stellarator symmetry and $n_{fp}$-symmetry in figure
\ref{fig:sensitivity_NEO_Fourier}.c-d.
Compared to the shape gradient for $\iota$, the shape gradient for $\epsilon_{eff}^{3/2}$ again has a large magnitude on the inboard side 
where the coil-plasma distance is small. However, this time there is also a significant magnitude on the outboard side of the type C coil, consistent
with the large magnitude of shape gradient near this toroidal angle ($\pi/3$) in figure \ref{fig:sensitivity_NEO_Fourier}.c-d.

The coil shape tolerance (\ref{eq:tolerance}) is displayed in figure \ref{fig:sensitivity_NEO_coils}.d. The acceptable variation 
is chosen to be $\Delta \epsilon_{eff}^{3/2}=\epsilon_{eff}^{3/2} / 2$, and again only the modular coils are considered.
In the main part of figure  \ref{fig:sensitivity_NEO_coils}.d, a weight $w_k$ scaling inversely with $|\vect{S}_k|$ is used
to focus the tight-tolerance regions to areas in which the shape and position of the coils
is most critical. Also shown on the same color scale is the uniform tolerance that gives the same bound on
$\delta \epsilon_{eff}^{3/2}$, resulting from the choice $w_k=1$.

%%%%%%%%%%%%%%%%%%%%%%%%%%%%%%%%%%%%%%%%%%%%%%%%%%%
%%%%%%%%%%%%%%%%%%%%%%%%%%%%%%%%%%%%%%%%%%%%%%%%%%%
%%%%%%%%%%%%%%%%%%%%%%%%%%%%%%%%%%%%%%%%%%%%%%%%%%%

\section{Cases in which the shape gradient does not exist}

% For details of calculations in this section, see 20171108-01 Example where sensitivity map does not exist.docx

Now consider a quantity $f$ that is not coordinate independent, that is, $f$ depends on the particular
choice of coordinates used to parameterize the plasma boundary shape.
This $f$ will change under perturbations of $\vect{r}(\theta,\zeta)$
that are tangential rather than normal to the surface, and hence perturbations
to $f$ cannot be expressed in the form 
(\ref{eq:Sdef3D}). If one attempts to compute a shape gradient  for $f$  using the machinery of section 
\ref{sec:finite_difference}, one will find that $\partial f/\partial\vect{p}$ will not be in the column space of
$\vect{D}$, even approximately.
Equivalently, the last half of the elements of $\vect{U}^T \, \partial f/\partial\vect{p}$ or $\vect{Q}^T \, \partial f/\partial\vect{p}$ will have a substantial amplitude. In this way, the method of section \ref{sec:finite_difference} can detect whether or not
a shape gradient exists, given the derivatives $\partial f/\partial p_j$ computed by
STELLOPT or some other method.

\begin{figure}[h!]
\includegraphics[width=3.5in]{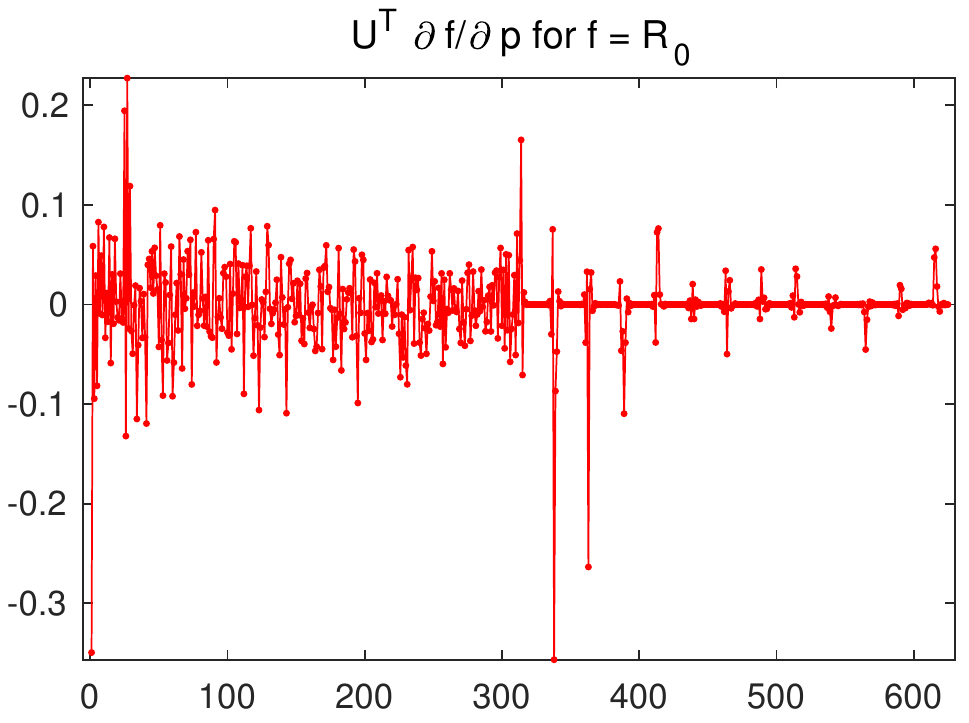}
\caption{(Color online)
Elements of the (dimensionless) vector $\vect{U}^{T} \, \partial R_0/\partial\vect{p}$, where $R_0$ is defined in (\ref{eq:R0}).
 The second half of the vector elements are not all negligible compared to the first half of the elements,
indicating that a shape gradient for $R_0$ does not exist.
\label{fig:R}}
\end{figure}

As an example, consider the following quantity, an averaged major radius:
\begin{equation}
R_0 = \frac{1}{(2\pi)^2}\int_0^{2\pi}d\theta \int_0^{2\pi}d\zeta \; R = R_{m=0,n=0}^c.
\label{eq:R0}
\end{equation}
A given surface can have different values of $R_0$ depending on the choice of $\theta$ used to parameterize the surface, and so no shape gradient for $R_0$ exists. Figure \ref{fig:R} shows
the elements of $\vect{U}^T \, \partial f/\partial\vect{p}$ computed for this figure of merit for NCSX (using VMEC's $\theta$).
For this example, poloidal and toroidal mode numbers up to 12
were retained in $p_j$; other choices yield qualitatively
similar results. 
The second half of the entries of $\vect{U}^T \, \partial f/\partial\vect{p}$ are not all small compared to the first
half of the entries, indicating that $\partial f/\partial\vect{p}$ is not in the column
space of $\vect{D}$, and hence the representation (\ref{eq:Sdef3D}) is not valid for $R_0$.

A shape gradient does however exist for any alternative effective major radius 
that is coordinate-independent, such as
$R_{\mathrm{eff}} = A^{-1} \int d^2a\; R$.

%%%%%%%%%%%%%%%%%%%%%%%%%%%%%%%%%%%%%%%%%%%%%%%%
%%%%%%%%%%%%%%%%%%%%%%%%%%%%%%%%%%%%%%%%%%%%%%%%

\section{Magnetic sensitivity}
\label{sec:magnetic}

By relating the displacement of the surface to magnetic field perturbations,
the magnetic sensitivity  $S_B$ in (\ref{eq:SB}) can be computed from the
shape gradient  $S$. The magnetic sensitivity can inform where coils should be placed to control
a particular figure of merit, and to what extent errors in the magnetic field can be tolerated.

To derive the magnetic sensitivity representation, we first write the total magnetic field as 
$\vect{B} = \vect{B}_0 + \delta \vect{B}$ where $\vect{B}_0$ is the field of the unperturbed
configuration. Similarly, we write $\psi = \psi_0 + \delta\psi$, where $\psi$ is any flux surface label coordinate
for the perturbed state, and $\psi_0$ is the analogous quantity for the unperturbed configuration.
The first-order terms in $\vect{B} \cdot\nabla\psi=0$ then give
\begin{equation}
\vect{B}_0 \cdot\nabla\delta\psi + \delta\vect{B}\cdot\nabla\psi_0=0.
\label{eq:BPerturbation1}
\end{equation}
Next, observe
\begin{equation}
0 = d\psi = \delta\psi + \delta\vect{r}\cdot\nabla\psi_0,
\label{eq:BPerturbation2}
\end{equation}
where $d\psi$ is the Lagrangian change in $\psi$ along the plasma boundary as it is deformed.
Combining (\ref{eq:BPerturbation1})-(\ref{eq:BPerturbation2}),
\begin{equation}
\vect{B}_0 \cdot\nabla ( \delta\vect{r}\cdot\nabla\psi_0) = \delta\vect{B}\cdot\nabla\psi_0.
\label{eq:BPerturbation3}
\end{equation}
From this result, we observe that the component of $\delta\vect{B}$ tangent to the unperturbed surface
causes no displacement; only the normal component matters. Also note that if the rotational transform $\iota$ on the surface is irrational, 
the magnetic differential equation (\ref{eq:BPerturbation3})
has a solvability condition $\left< \delta\vect{B}\cdot\nabla\psi_0\right>=0$
where 
angle brackets denote a flux surface average:
$\left< Q \right> = (1/V') \int_0^{2\pi}d\theta \int_0^{2\pi}d\zeta \sqrt{g} Q$ where $V' =\int_0^{2\pi}d\theta \int_0^{2\pi}d\zeta  \sqrt{g}$ for any quantity $Q$, and 
\begin{equation}
\sqrt{g} = \frac{\partial\vect{r}}{\partial\psi_0} \cdot \frac{\partial\vect{r}}{\partial\theta} \times
\frac{\partial\vect{r}}{\partial\zeta}
\label{eq:sqrt_g}
\end{equation}
is the Jacobian.
It can be shown that this solvability condition is always satisfied due to the absence of magnetic monopoles.
If on the other hand $\iota$ is rational, the solvability condition is obtained by integrating (\ref{eq:BPerturbation3})
over a closed field line rather than a surface, and a resonant magnetic field perturbation can violate this solvability condition,
associated with the fact that $\psi$ cannot be defined if the perturbed boundary becomes a magnetic island.
We will not consider this resonant situation further.
Also note that for a given $\delta\vect{B}$, (\ref{eq:BPerturbation3}) only determines $\delta\vect{r}\cdot\nabla\psi_0$
up to a constant. This constant homomgeneous solution exists because different 
magnetic surfaces could be identified as the boundary surface.

Equation (\ref{eq:BPerturbation3}) can also be derived from ideal magnetohydrodynamics (MHD).
In this plasma model, the linearized induction equation is $\delta \vect{B} = \nabla\times(\delta\vect{r}\times\vect{B}_0)$.
Applying $\cdot\nabla\psi_0$, the right-hand side can be manipulated using vector identities to yield (\ref{eq:BPerturbation3}).

Next, we define $S_B$ as the solution of
\begin{equation}
\vect{B}_0\cdot\nabla S_B = \left<S\right> - S,
\label{eq:SB_def}
\end{equation}
where again we assume the boundary $\iota$ is irrational.
This equation only determines $S_B$ up to a constant, and we will show later that this constant is unimportant.
We then substitute (\ref{eq:SB_def}) into (\ref{eq:Sdef3D}),
using the fact that area integrals can be written 
$\int d^2a (\ldots) = \int_0^{2\pi} d\theta \int_0^{2\pi}d\zeta |\sqrt{g}| |\nabla\psi_0| (\ldots)$,
obtaining
\begin{equation}
\delta f = \left< S\right> \int d^2a  \, \delta\vect{r} \cdot \vect{n}
- \int_0^{2\pi}d\theta \int_0^{2\pi}d\zeta |\sqrt{g}| \left( \delta\vect{r}\cdot\nabla\psi_0\right) \vect{B}_0\cdot\nabla S_B.
\end{equation}
Integrating the last term by parts, and substituting (\ref{eq:BPerturbation3}),
we obtain (\ref{eq:SB}),
where $\delta V =  \int d^2a  \, \delta\vect{r} \cdot \vect{n}$ is the perturbation to the volume.
Aside from the $\left< S \right>$ term, this result resembles the shape gradient  form (\ref{eq:Sdef3D}) but with the displacement $\delta\vect{r}$ replaced by the magnetic perturbation $\delta\vect{B}$.
However, as noted following (\ref{eq:sqrt_g}), a given $\delta\vect{B}$ does not determine the constant part of $\delta\vect{r} \cdot\vect{n}$, and hence the $\left< S\right>$ term in (\ref{eq:SB}) is necessary.
Also note that if we had not included $\left< S\right>$ in (\ref{eq:SB_def}), the solvability condition
of (\ref{eq:SB_def}) would generally not be satisfied. 

Note that shifting $S_B$ by a constant does not alter $\delta f$ in (\ref{eq:SB}),
since $\int d^2a \, \delta\vect{B}\cdot\vect{n}=0$ by the absence of magnetic monopoles. 
Hence, the homogeneous solution of (\ref{eq:SB_def}) 
is unimportant.
For results shown below, we choose the constant so $\left< S_B\right>=0$. 

In summary, any time a shape gradient $S$ exists and the boundary $\iota$ is irrational,
the magnetic sensitivity  representation (\ref{eq:SB}) also exists,
and the magnetic sensitivity $S_B$ can be computed from $S$ by (\ref{eq:SB_def}).

Note that $\delta\vect{B}$ throughout this analysis is a sum of contributions from currents
outside the plasma plus any  response from currents inside the plasma. In the limit of small plasma $\beta$,
the plasma response can be neglected, so then $\delta\vect{B}$ can be identified with the external perturbation.

\begin{figure}[h!]
\includegraphics[width=3.5in]{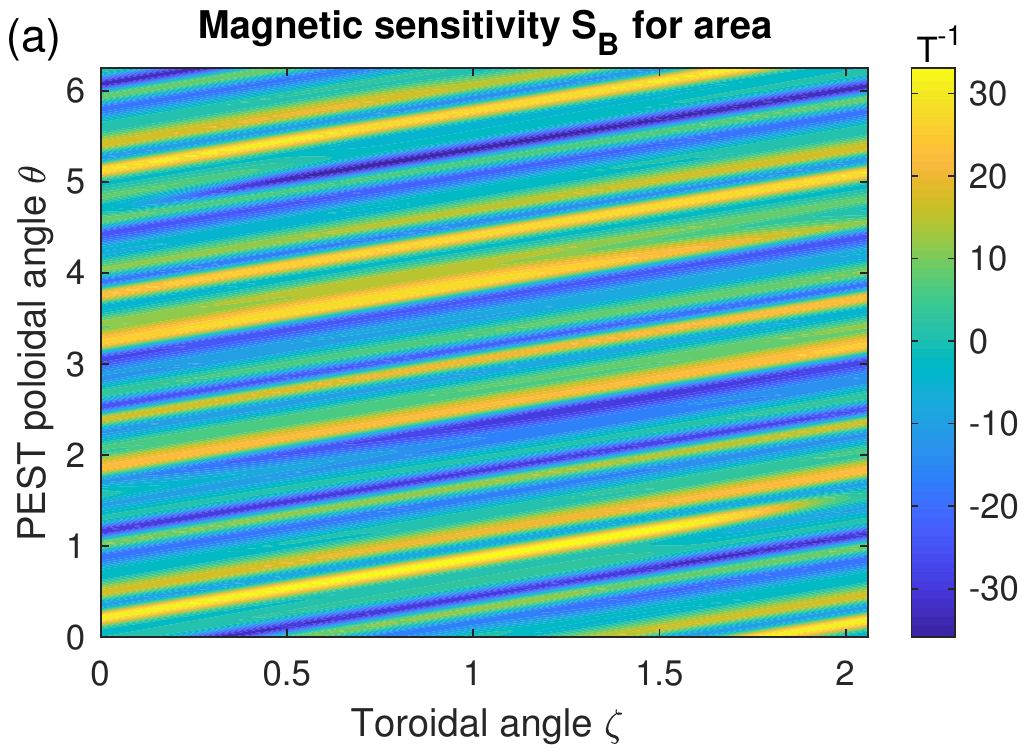}
\includegraphics[width=3.5in]{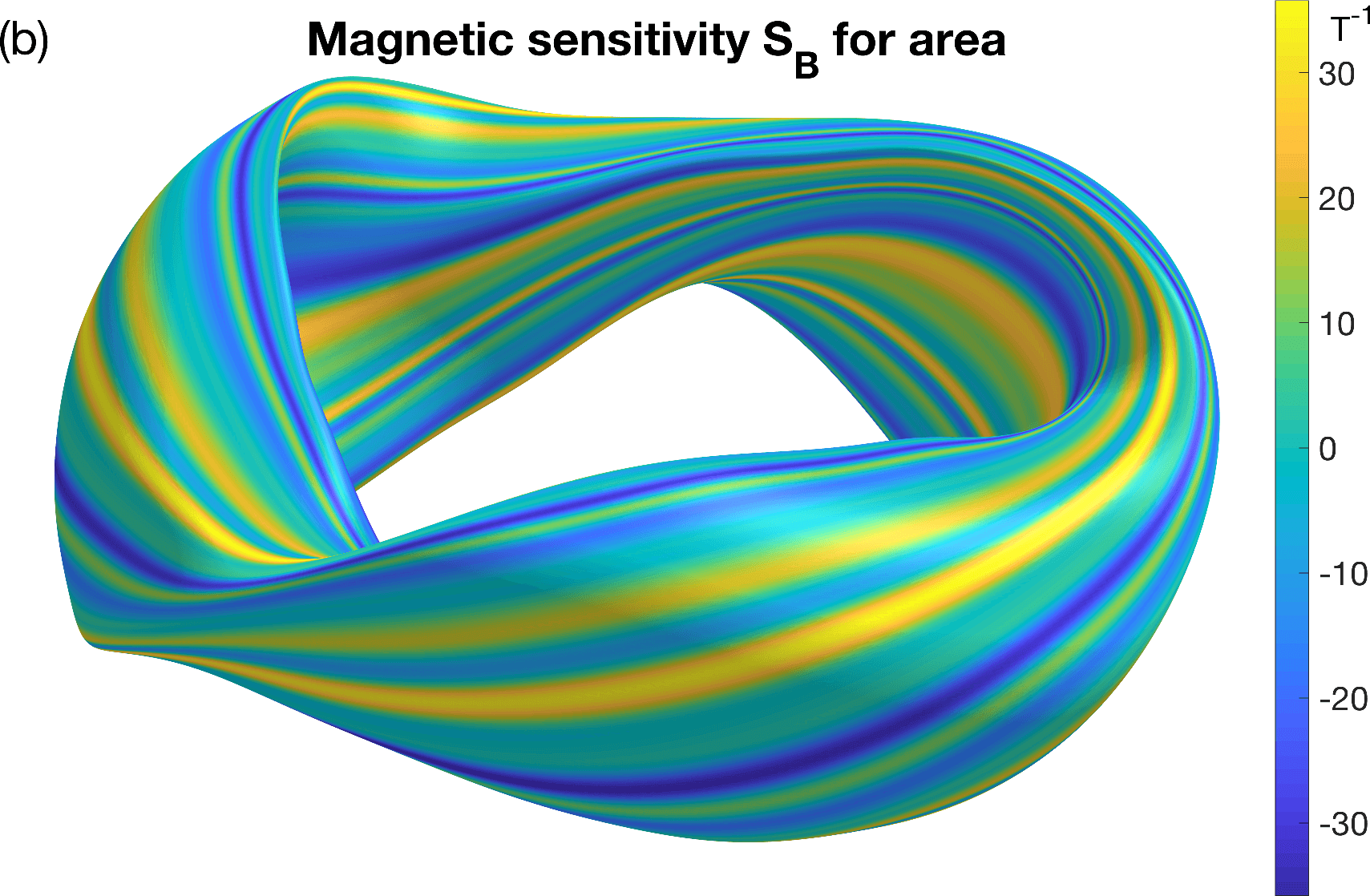}
\\
\vspace{0.2in}
\includegraphics[width=3.5in]{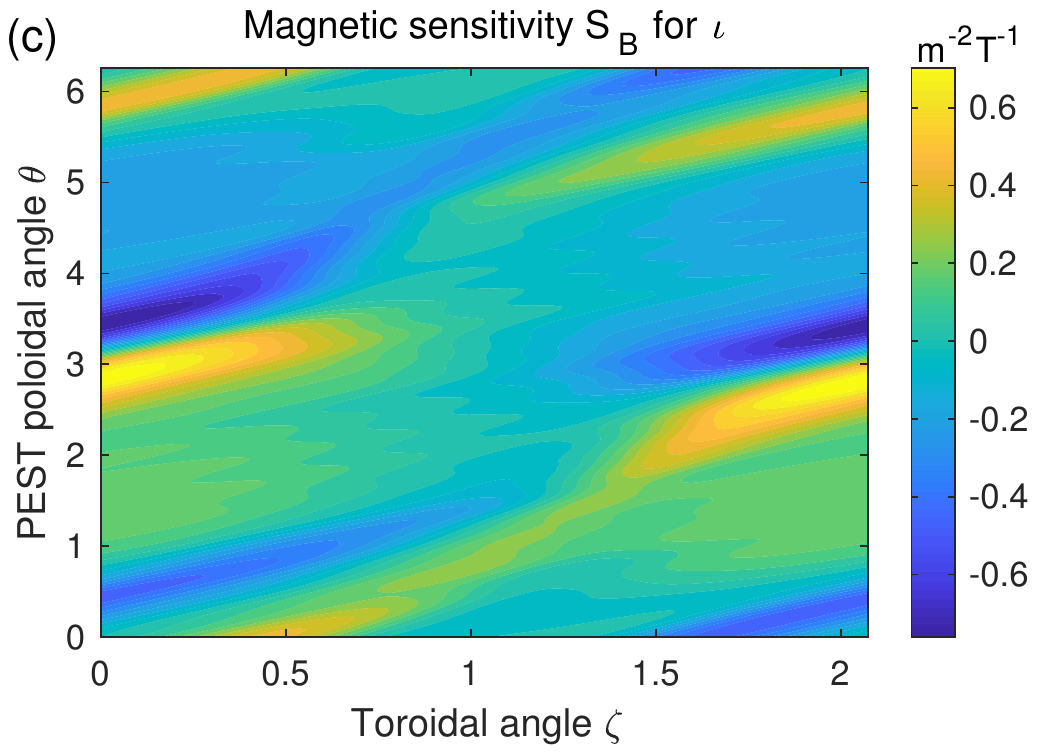}
\includegraphics[width=3.5in]{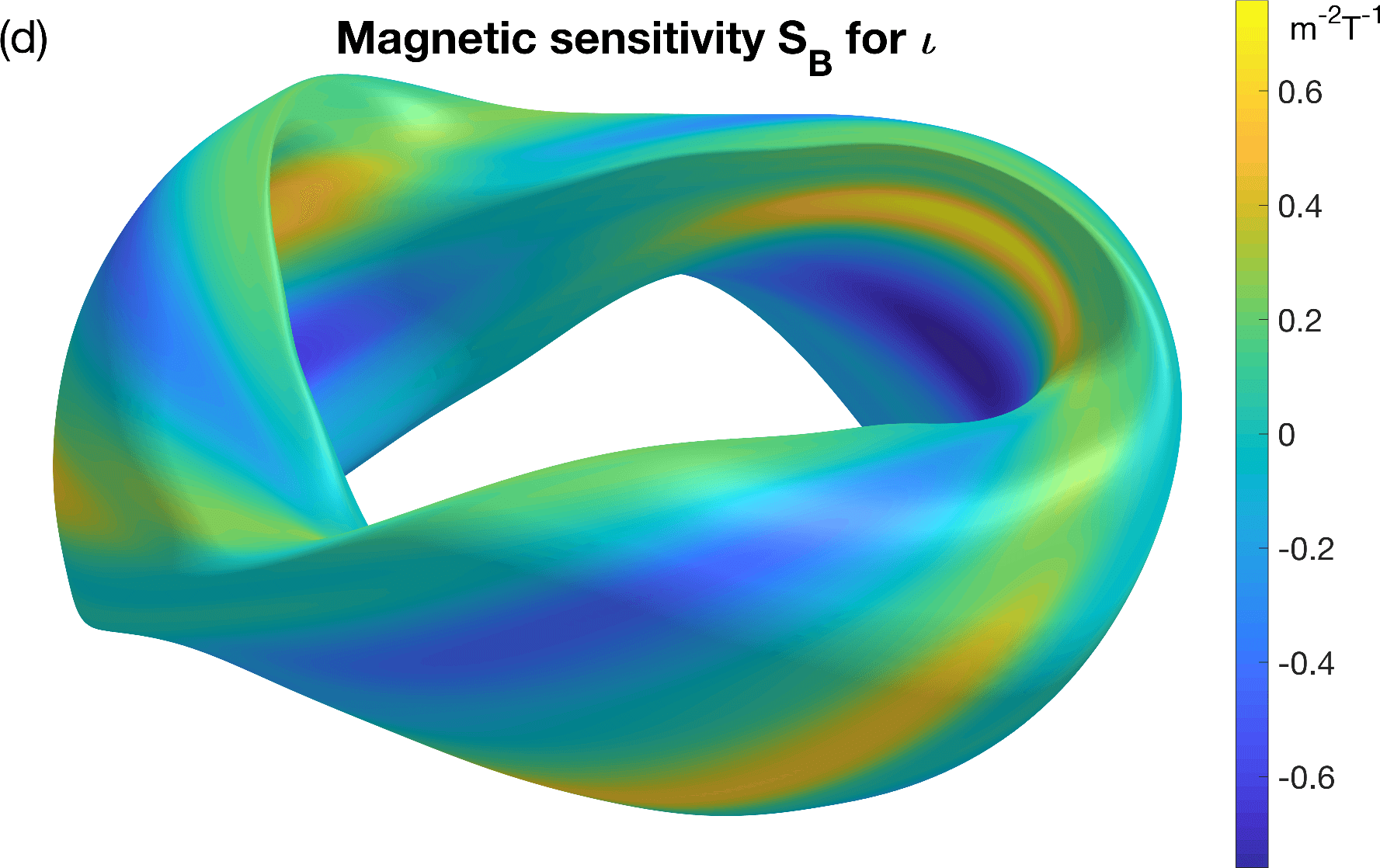}
\\
\vspace{0.2in}
\includegraphics[width=3.5in]{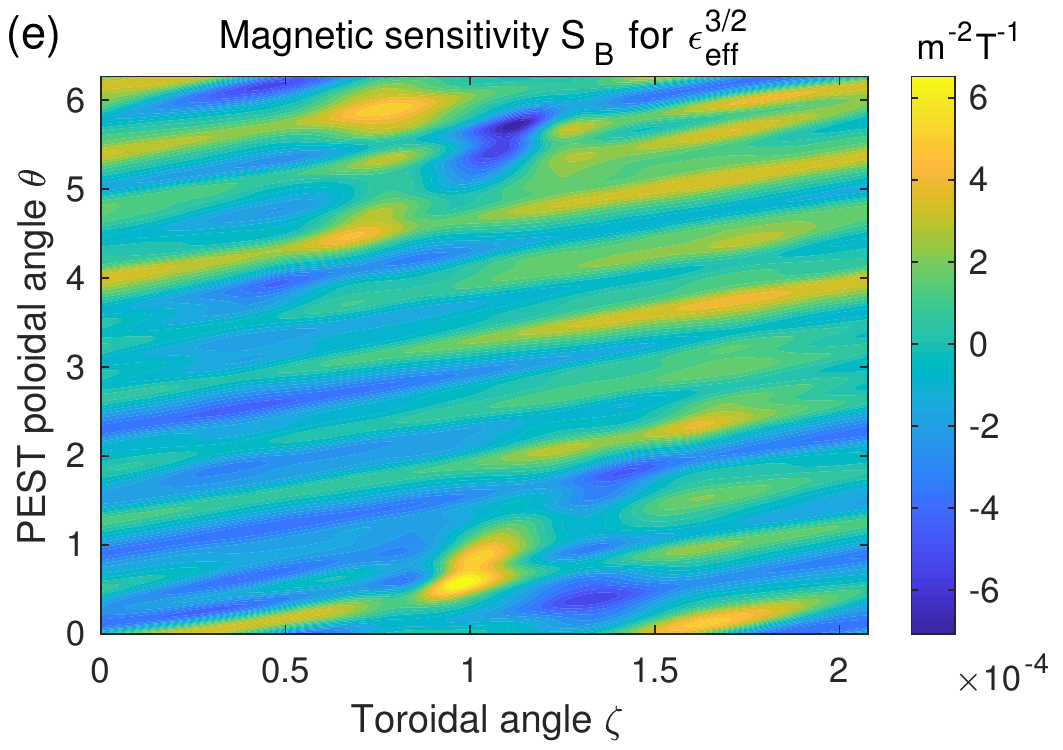}
\includegraphics[width=3.5in]{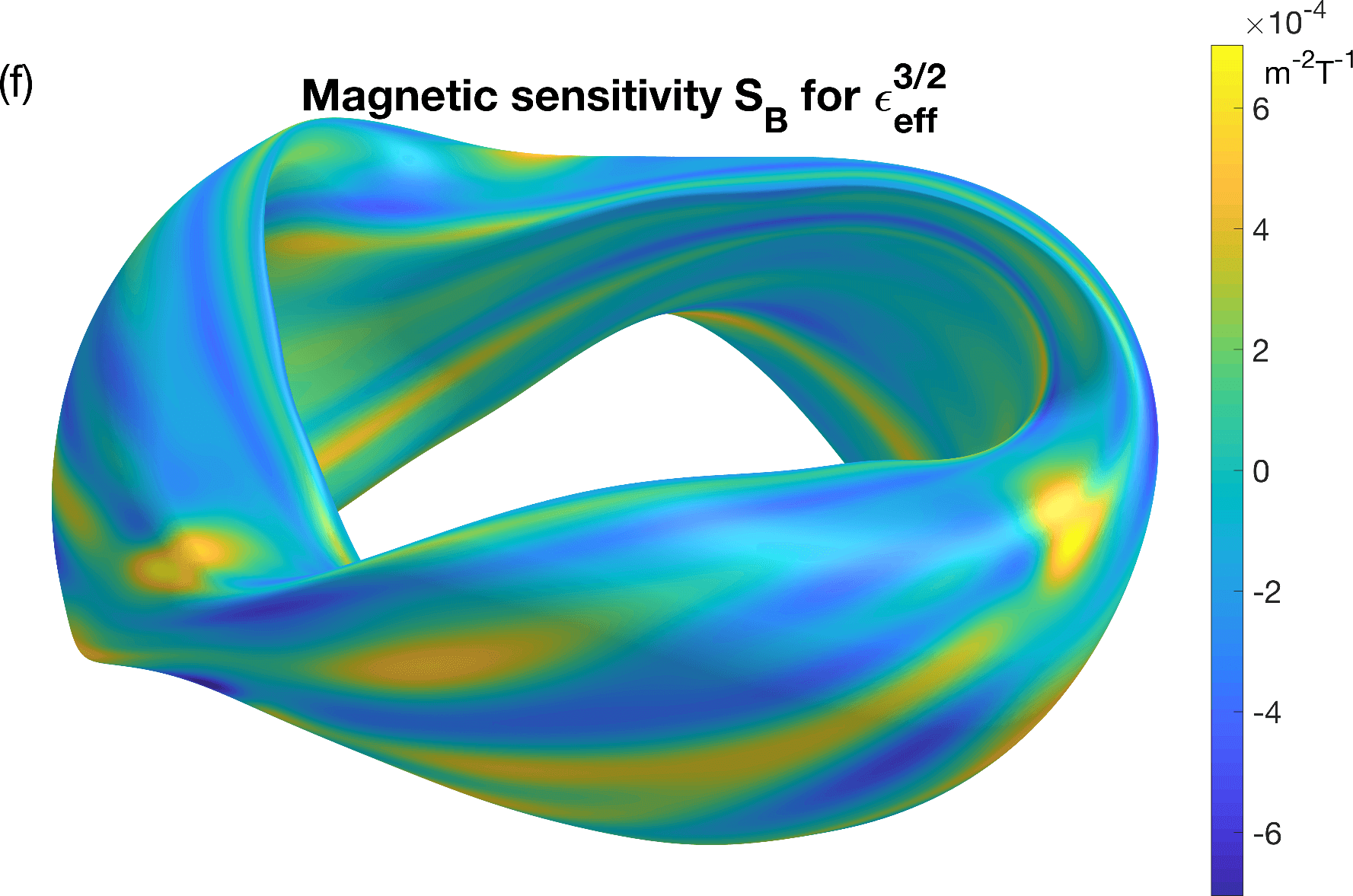}
\caption{(Color online)
Magnetic sensitivity, computed using the methods of section \ref{sec:magnetic}.
\label{fig:magnetic}}
\end{figure}

Figure \ref{fig:magnetic} shows the magnetic sensitivity 
computed for the examples considered previously.
To produce these figures, equation  (\ref{eq:SB_def}) is solved in Fourier space using straight-field-line (PEST) coordinates.
The regions in these figures where the magnitude of sensitivity is large would be ideal locations for coils that 
control the relevant figure of merit while having less effect on other physics properties.

Just as a tolerance for shape deformations was derived from the shape gradient in section \ref{sec:tolerance},
a local tolerance for magnetic field errors can be derived from the magnetic sensitivity. 
This local magnetic tolerance can be useful for identifying if 
potential sources of error fields, such as small amounts of magnetic material
or current leads, can be tolerated, and where they should be located.
First, for any chosen figure of merit $f$, an acceptable overall $\Delta f>0$ is identified. Then we
define the local magnetic tolerance $T_B$ at each point on the plasma boundary by
\begin{equation}
T_B(\theta,\zeta) =  \frac{W(\theta,\zeta)\, \Delta f }{ \int d^2a' \,W(\theta',\zeta') |S_B(\theta',\zeta')|},
\label{eq:magnetic_tolerance}
\end{equation}
where $W$ is any non-negative weight.
If the normal component of magnetic field error is everywhere smaller than $T_B$,
i.e. $| \delta \vect{B} \cdot\vect{n} | \le T_B$,
then $\delta f$ 
can be guaranteed
to be no larger in magnitude than $\Delta f$:
\begin{equation}
|\delta f| \le \int d^2a |S_B| |\delta \vect{B}\cdot\vect{n}|
\le \int d^2a |S_B| T_B
\le \Delta f.
\end{equation}
In the line above we have set $\delta V=0$ in (\ref{eq:SB}), i.e. we choose to define the perturbed plasma boundary to be the perturbed
flux surface with equal volume to the unperturbed case.
The choice $W = 1$ makes the tolerance spatially uniform, $T_B = \Delta f / \left( \int d^2a' |S_B(\theta',\zeta')| \right)$.
Or, if the weight $W$ is chosen to scale inversely with $|S_B|$ in some way,
such as $W = |S_B|^{-\alpha}$ for some $\alpha \ge 0$, the tolerance 
can be relaxed over a majority of the plasma surface in exchange
for a tighter tolerance in a few selected locations where the plasma is particularly sensitive.
Other choices for $W$ could be
made to allocate more of the tolerance to specified regions, which might be useful if
a source of magnetic field error is known to be localized.
As with the coil shape tolerance (\ref{eq:tolerance}), (\ref{eq:magnetic_tolerance}) is conservative
in the sense that it is a limit on the largest possible $|\delta f|$, rather than a limit on
the expectation value of $|\delta f|$ from some anticipated distribution of errors.

\begin{figure}[h!]
\includegraphics[width=3.5in]{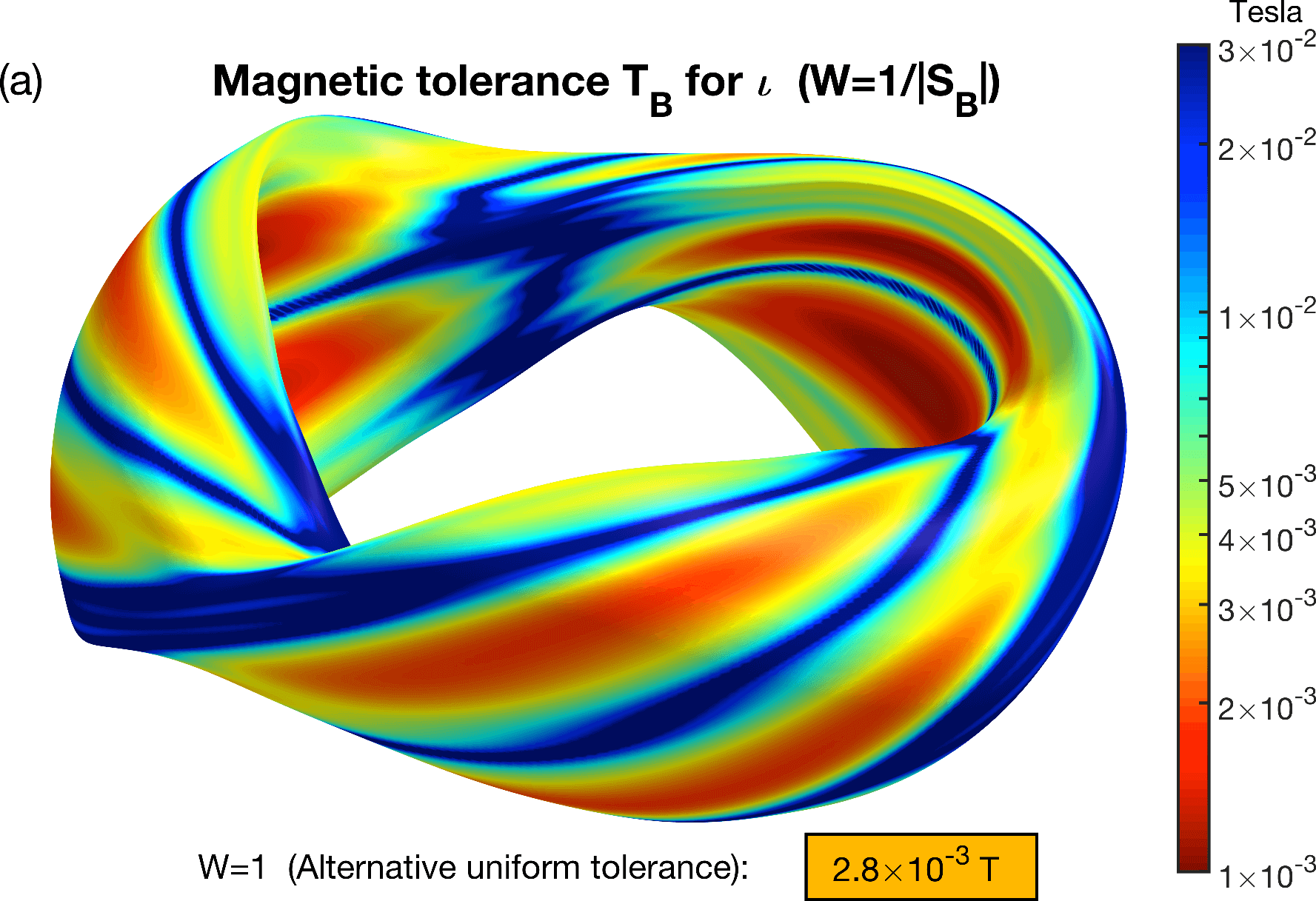}
\includegraphics[width=3.5in]{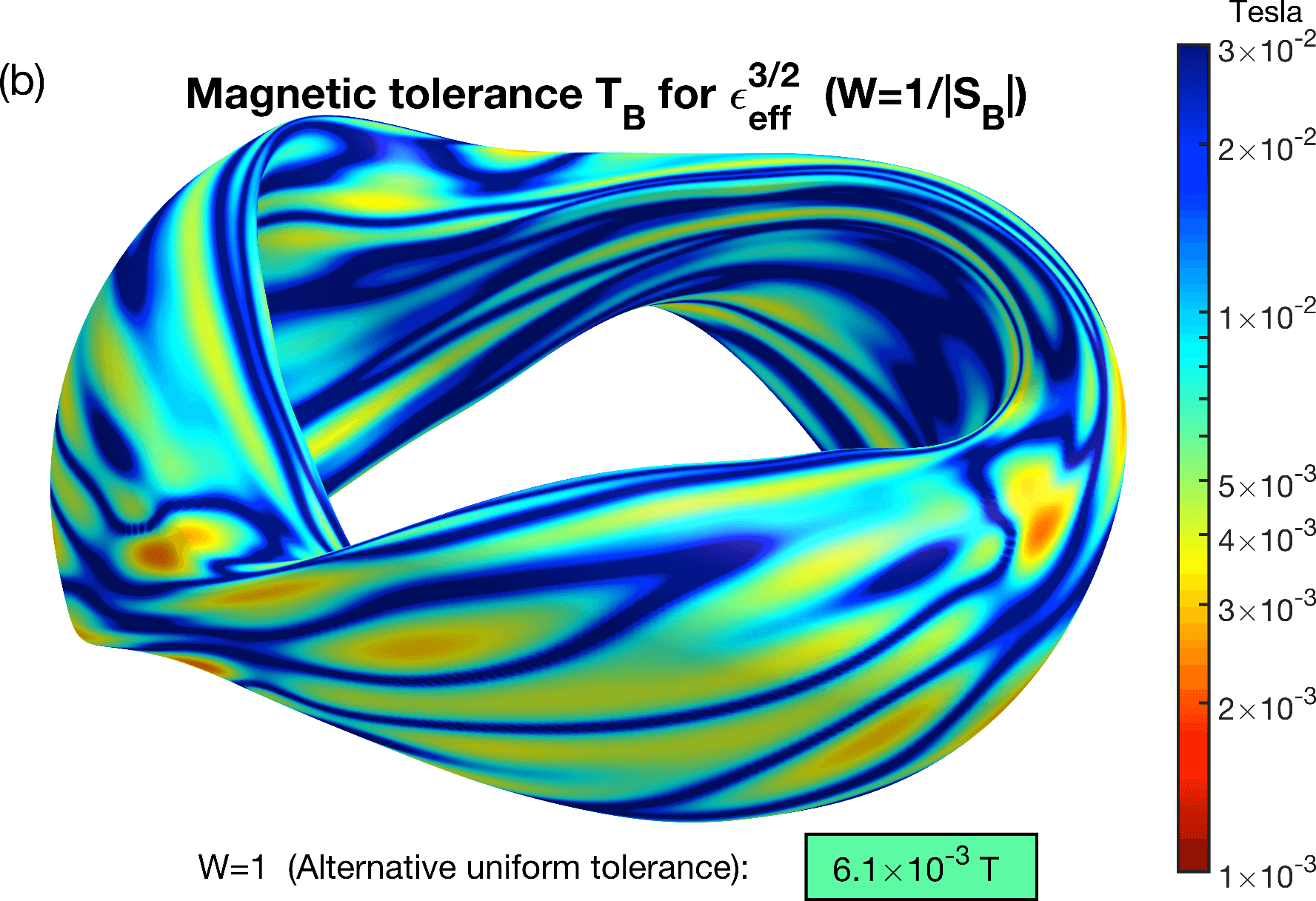}
\caption{(Color online)
The magnetic tolerance (\ref{eq:magnetic_tolerance}) for (a) the half-radius rotational transform and (b) the half-radius neoclassical
transport, computed using acceptable variations $\Delta \iota = 0.02$ and $\Delta \epsilon_{eff}^{3/2} = \epsilon_{eff}^{3/2}/2$.
In the 3D figures, a nonuniform spatial weighting is used to highlight the regions that are most sensitive, while the uniform tolerance
that yields the same bound on $\Delta f$ is shown below on the same color scale.
\label{fig:magnetic_tolerance}}
\end{figure}

Figure \ref{fig:magnetic_tolerance} shows the magnetic tolerances computed by applying (\ref{eq:magnetic_tolerance})
to the magnetic sensitivities shown in figure \ref{fig:magnetic}.c-f.
The allowable variations are the same as for the coil tolerances in figures \ref{fig:sensitivity_iota_coils}.c and 
\ref{fig:sensitivity_NEO_coils}.d: $\Delta \iota = 0.02$ and $\Delta \epsilon_{eff}^{3/2} = \epsilon_{eff}^{3/2}/2$.
In the 3D part of the figures, the weight is chosen to be $W=1/|S_B|$, yielding a `tolerance map' that highlights
the regions where the plasma is most sensitive. The uniform tolerance obtained using $W=1$ is also displayed, 
and it is between the upper and lower bounds of the nonuniform tolerance.

%%%%%%%%%%%%%%%%%%%%%%%%%%%%%%%%%%%%%%%%%%%%%%%%%%%
%%%%%%%%%%%%%%%%%%%%%%%%%%%%%%%%%%%%%%%%%%%%%%%%%%%
%%%%%%%%%%%%%%%%%%%%%%%%%%%%%%%%%%%%%%%%%%%%%%%%%%%

\section{Conclusions}

In this paper we have pointed out that derivatives of many figures of merit $f$
with respect to the shape of a plasma or coil can be represented
not only by the parameter derivatives $\partial f/\partial p_j$,
but also by the shape gradient representations (\ref{eq:Sdef3D})-(\ref{eq:Sdef_3Dcurve}),
and the shape gradient representations have several advantages.
These representations provide information that is spatially local,
which is important for  engineering considerations,
and the shape gradients are also independent of whichever parameterization
is chosen for the shape.
Most existing stellarator physics codes do not provide derivative information,
but we have demonstrated how the shape gradients can nonetheless be computed from any `forward code'
such as those in the STELLOPT code suite.
First, the parameter derivatives $\partial f/\partial p_j$ can be computed from the forward code using
finite differences, and then the shape gradient can be computed
from the parameter derivatives by solving the linear system (\ref{eq:Sdef3Dcurves_alt}) for coils
or (\ref{eq:Sdef3D_alt}) for surfaces.
Using this procedure, shape gradients for several interesting stellarator figures of merit have been
illustrated in figures \ref{fig:sensitivity_iota_Fourier}-\ref{fig:sensitivity_NEO_coils}.
These shape gradients are more illuminating in several ways than 
the equivalent parameter derivatives in figures \ref{fig:sensitivity_iota_Fourier}.a and \ref{fig:sensitivity_NEO_Fourier}.a.
Once the shape gradient is found for a toroidal surface, the magnetic sensitivity $S_B$
can be computed by solving (\ref{eq:SB_def}), enabling variations with respect to local
magnetic field perturbations to be represented in the form (\ref{eq:SB}).
The magnetic sensitivity, as shown in figure \ref{fig:magnetic}, can help inform where to locate trim coils.
Both the shape gradient for coils and the magnetic sensitivity for surfaces can be converted
into local tolerances using (\ref{eq:tolerance}) and (\ref{eq:magnetic_tolerance}).
The shape tolerance, as shown in figures \ref{fig:sensitivity_iota_coils}.c and \ref{fig:sensitivity_NEO_coils}.d,
informs how accurately the coils must be built and positioned, and where
the coils should be most rigidly supported.
The magnetic tolerance, as shown in figure \ref{fig:magnetic_tolerance}, indicates 
the extent to which stray field from magnetic material and current leads is permissible, and where these sources
of error fields could best be located.
The shape gradient representation has been used previously for shape optimization of objects
interacting with neutral fluids, such as components of aircraft or automobiles,
and has been used to optimize the shapes of tokamak divertors.
Going forward, we expect this shape gradient representation will also be a powerful tool for optimizing plasma and coil shapes,
and for understanding their tolerances.

\begin{acknowledgments}
We are especially grateful to Thomas Antonsen for bringing the shape gradient representation to our attention,
and for making suggestions that improved the manuscript.
Dr Carsten Othmer was gracious in welcoming 
 the use of figure \ref{fig:car}.
We also acknowledge illuminating conversations about this work with 
Stuart Hudson,
Neil Pomphrey,
Georg Stadler,
and Thomas Sunn Pedersen,
and for assistance with the STELLOPT code suite from Sam Lazerson.
This work was supported by the
U.S. Department of Energy, Office of Science, Office of Fusion Energy Science,
under Award Number DE-FG02-93ER54197.
\end{acknowledgments}

\appendix

\section{Shape gradient for area integrals}
\label{appendix:area}

Here we give a derivation of (\ref{eq:area_integrals}).
Perturbing $\vect{r}(\theta,\zeta)$ in (\ref{eq:A})-(\ref{eq:bigN}), we find
\begin{equation}
\delta A = \int_0^{2\pi} d\theta \int_0^{2\pi}d\zeta 
\left( Q\frac{\partial \delta \vect{r}}{\partial\zeta}\cdot \frac{\partial\vect{r}}{\partial\theta} \times\vect{n}
+ Q\frac{\partial\delta \vect{r}}{\partial\theta} \cdot\vect{n} \times \frac{\partial\vect{r}}{\partial \zeta} 
+N \delta\vect{r}\cdot\nabla Q \right).
\end{equation}
Integrating by parts to remove the $\theta$ and $\zeta$ derivatives from $\delta\vect{r}$,
\begin{equation}
\delta A = \int d^2 a \left[ \delta\vect{r} \cdot (Q\vect{T} + \nabla Q)
+ \frac{1}{N} \delta \vect{r}\times\vect{n} \cdot \left( \frac{\partial\vect{r}}{\partial \theta} \frac{\partial Q}{\partial\zeta} 
-\frac{\partial\vect{r}}{\partial\zeta} \frac{\partial Q}{\partial\theta} \right) \right],
\label{eq:deltaA}
\end{equation}
where
\begin{equation}
\vect{T} = \frac{1}{N} \left( \frac{\partial\vect{n}}{\partial\zeta} \times \frac{\partial\vect{r}}{\partial\theta}
+ \frac{\partial\vect{r}}{\partial\zeta} \times \frac{\partial\vect{n}}{\partial\theta} \right).
\label{eq:areaSensitivity}
\end{equation}
One can verify that $\vect{T}\cdot (\partial\vect{r}/\partial\theta) = 0$ and $\vect{T}\cdot (\partial\vect{r}/\partial\zeta) = 0$,
so $\vect{T}$ must be parallel to $\vect{N}$ and hence
$\vect{T}=\vect{T}\cdot\vect{n}\vect{n}$.
Expanding $\vect{n} = N^{-1}\vect{N}$ in (\ref{eq:areaSensitivity}) using
(\ref{eq:bigN}), 
one finds
\begin{equation}
\vect{T}\cdot\vect{n} 
= -\frac{LG+PE-2MF}{EG-F^2} = -2 H,
\label{eq:S_area_mean_curvature}
\end{equation}
where $E = (\partial\vect{r}/\partial\theta) \cdot (\partial\vect{r}/\partial\theta)$,
$F = (\partial\vect{r}/\partial\theta) \cdot (\partial\vect{r}/\partial\zeta)$, and 
$G = (\partial\vect{r}/\partial\zeta) \cdot (\partial\vect{r}/\partial\zeta)$
are the coefficients of the first fundamental form;
$L = \vect{n}\cdot (\partial^2\vect{r}/\partial \theta^2)$,
$M = \vect{n}\cdot (\partial^2\vect{r}/\partial \theta \partial\zeta)$, and
$P = \vect{n}\cdot (\partial^2\vect{r}/\partial \zeta^2)$
are the coefficients of the second fundamental form;
and $H$ is the mean curvature.
Recognizing the quantity in the last pair of parentheses in (\ref{eq:deltaA}) as $\vect{N}\times\nabla Q$, 
we finally obtain the form (\ref{eq:Sdef3D}) with a shape gradient given by (\ref{eq:area_integrals}).

%%%%%%%%%%%%%%%%%%%%%%%%%%%%%%%%%%%%%%%%%%%%%%%%%%%%%%%%%%%%%
%%%%%%%%%%%%%%%%%%%%%%%%%%%%%%%%%%%%%%%%%%%%%%%%%%%%%%%%%%%%%

\section{Collocation method for computing the shape gradient}
\label{sec:collocation}

For the method of computing the shape gradient in section \ref{sec:finite_difference},
$S$ was discretized using a finite Fourier expansion. Alternatively, $S$ could be represented
discretely by its values at the grid points in $\theta$ and $\zeta$ where $(\partial\vect{r}/\partial p_j)\cdot\vect{n}$
is evaluated. In this `collocation' approach, the matrix $\vect{D}$ of section \ref{sec:finite_difference}
is replaced by the matrix $\hat{\vect{D}}$ with elements
\begin{equation}
\hat{D}_{jq} = \Delta\theta \, \Delta\zeta \frac{\partial\vect{r}}{\partial p_j}\cdot \vect{N} 
\;\; \mbox{ evaluated at } (\theta_q,\zeta_q).
\end{equation}
As before, $q$ indexes a uniform tensor product grid in $\theta$ and $\zeta$ with spacing $\Delta\theta$ and $\Delta\zeta$.
Also, the vector $\vect{S}$ with elements $S_q$ is replaced by $\hat{\vect{S}}$, with elements $\hat{S}_q=S(\theta_q,\zeta_q)$. The rest of the discussion in section \ref{sec:finite_difference}
then applies. One can test whether the shape gradient representation exists by checking that $\partial f/\partial\vect{p}$ lies (approximately)
in the column space of $\hat{\vect{D}}$, and if it does, then
the shape gradient $\hat{\vect{S}}$ can be computing by applying the pseudo-inverse of $\hat{\vect{D}}$ to $\partial f/\partial \vect{p}$.

In contrast to the Fourier approach of section \ref{sec:finite_difference}, the results of the collocation method
are rather sensitive to the number of points in the $(\theta,\zeta)$ grid relative to the number of Fourier modes in $\partial f/\partial p_j$.
Also, if the $(\theta,\zeta)$ resolution is sufficiently high compared to the number of Fourier modes in $\partial f/\partial p_j$,
then $\hat{\vect{D}}$ will be rank-deficient, and so its singular values below some threshold must be treated as if they were zero
in the pseudo-inverse. Without a singular value threshold, the computed shape gradient becomes dominated by numerical noise.
Due to these issues, the collocation approach seems less robust than the Fourier approach.
For the examples considered in this paper, we find it works well to use three times as many grid points
in $\theta$ as the largest $m$ mode in $\partial f/\partial p_j$, use three times as many grid points
in $\zeta$ as the largest $n$ mode in $\partial f/\partial p_j$, and use a singular value threshold of $\sim 0.05 \times$ the largest singular value.

\begin{figure}[h!]
\includegraphics[width=3.5in]{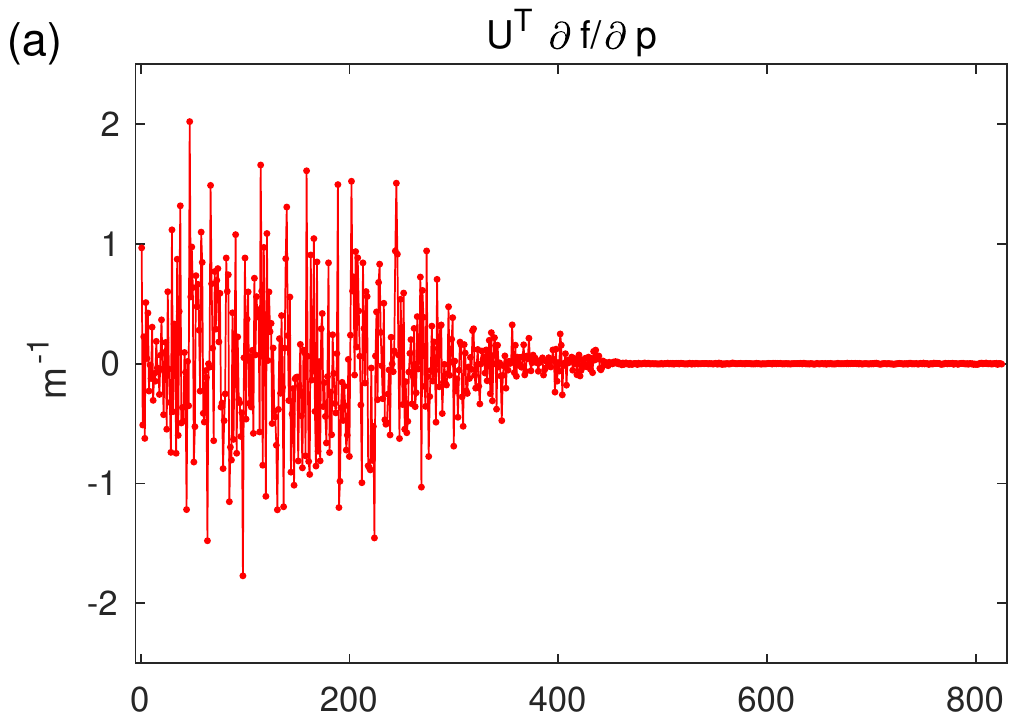}
\includegraphics[width=3.5in]{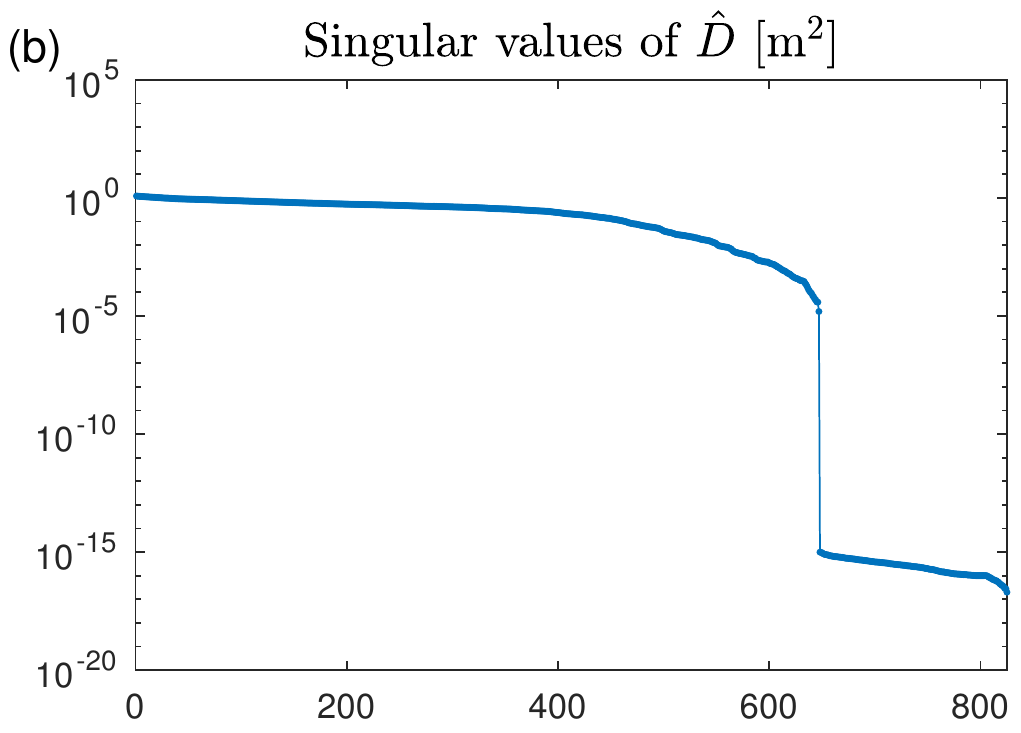}
\vspace{0.2in}
\includegraphics[width=3.5in]{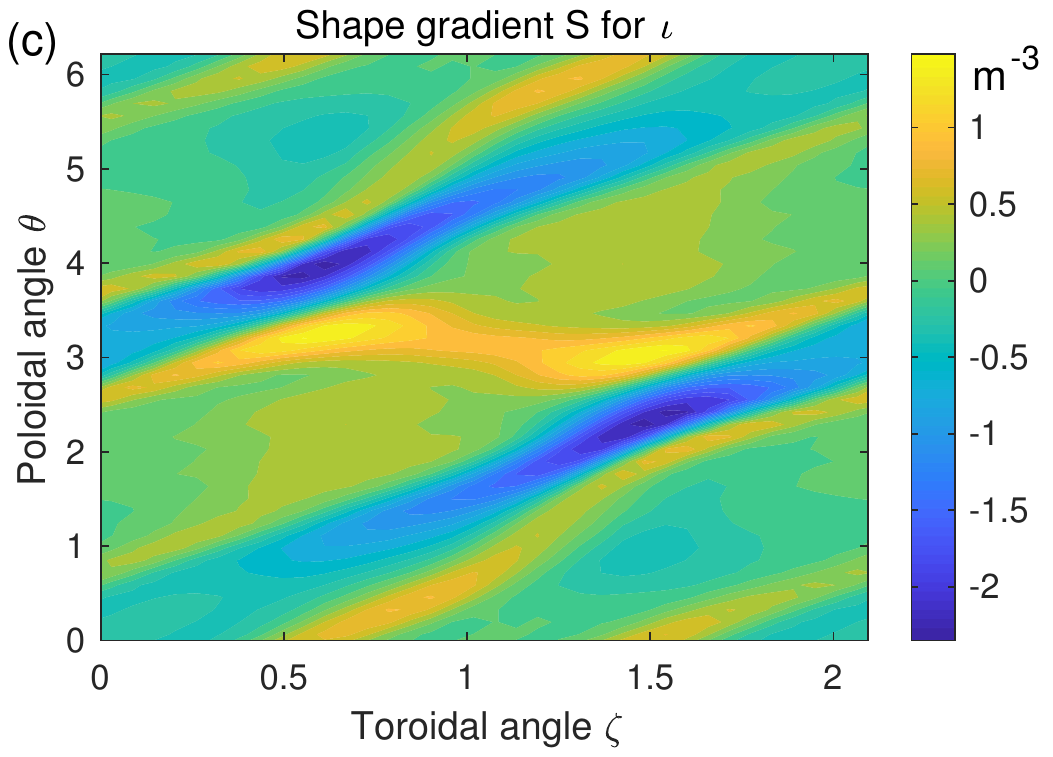}
\caption{(Color online)
Computation of the shape gradient for the half-radius rotational transform $\iota$ using
the collocation approach  of Appendix \ref{sec:collocation}.
(a) Projection of $\partial f/\partial \vect{p}$ onto the left singular vectors of $\hat{\vect{D}}$. 
(b) Singular values of $\hat{\vect{D}}$. The small singular values are all associated with negligible values
in (a), confirming that changes to $\iota$ depend only on the normal component of displacement,
not on the tangential component.
(c) The resulting shape gradient for $\iota$ is nearly indistinguishable from figure \ref{fig:sensitivity_iota_Fourier}.c.
\label{fig:collocation}}
\end{figure}

This collocation method for computing the shape gradient is illustrated in figure \ref{fig:collocation}.
Here, the rotational transform example of section \ref{sec:iota} and figure \ref{fig:sensitivity_iota_Fourier} is considered.
Due to stellarator-symmetry and $n_{fp}$-symmetry, the $(\theta,\zeta)$ grid need only extend over half of one field period.
Figure \ref{fig:collocation}.a shows $\vect{U}^T \partial f/\partial \vect{p}$,
where $\vect{U}$ is the matrix of left singular vectors of $\hat{\vect{D}}$.
Figure \ref{fig:collocation}.b shows the singular values of $\hat{\vect{D}}$.
It can be seen that the small singular values correspond to negligible entries in $\vect{U}^T \partial f/\partial \vect{p}$,
and so $\partial f/\partial \vect{p}$ does lie in the column space of $\hat{\vect{D}}$. Hence
a shape gradient does exist.
Figure \ref{fig:collocation}.c shows the final shape gradient. The result is nearly indistinguishable from
figure figure \ref{fig:sensitivity_iota_Fourier}.d, demonstrating that the Fourier and collocation methods yield essentially the same results.
The same is true for the area and $\epsilon_{eff}^{3/2}$ examples.

\section{Symmetries}
\label{sec:symmetry}

Stellarator magnetic surfaces often have two discrete symmetries. The first, `$n_{fp}$ symmetry', is a discrete rotational symmetry: the system is unchanged under toroidal rotation $\zeta \to \zeta + 2\pi/n_{fp}$. Here, the integer $n_{fp}$ denotes the number of field periods (5 for W7-X, 3 for NCSX). The second symmetry is stellarator symmetry: $R(\theta,\zeta)=R(-\theta,-\zeta)$ and $Z(\theta,\zeta)=-Z(-\theta,-\zeta)$.

Even if a toroidal surface has $n_{fp}$ symmetry and stellarator symmetry, there is no general rule that the shape gradient $S$
must have the same symmetries for all shape functionals. For example, consider the functional $f = \int d^3r\; (X+Z)$, where $X$ and $Z$ are Cartesian coordinates. The associated shape gradient, using the result of section \ref{sec:volume_integrals}, is $S=X+Z$, which lacks both symmetries.
However, for many figures of merit of physical interest, $S$ does posses one or both symmetries.

In the remainder of this section, we prove that for a stellarator-symmetric surface shape, $S$ is stellarator-symmetric
if and only if $\partial f/\partial R_{m,n}^s=0$ and $\partial f/\partial Z_{m,n}^c=0$ for all $m$ and $n$.
We also prove that for a $n_{fp}$-symmetric surface shape, $S$ is $n_{fp}$-symmetric
if and only if $\partial f/\partial R_{m,n}^c$, $\partial f/\partial R_{m,n}^s$, $\partial f/\partial Z_{m,n}^c$, and $\partial f/\partial Z_{m,n}^s$
all vanish when $n$ is not a multiple of $n_{fp}$. In these statements, $f$ can be any shape functional.
Using these results, one can determine whether $S$ has $n_{fp}$ symmetry and/or stellarator symmetry based on the parameter derivatives.
If $S$ is found to have one or both symmetries, the discretized linear systems described in section \ref{sec:finite_difference} and appendix \ref{sec:collocation} can be reduced in size accordingly.

Let $(\vect{e}_R,\vect{e}_\zeta,\vect{e}_Z)$ be the usual right-handed set of unit vectors for cylindrical coordinates. 
Using $\vect{r} = R \vect{e}_R + Z \vect{e}_Z$, $d \vect{e}_R/d\zeta = \vect{e}_\zeta$, and 
(\ref{eq:bigN}), then
\begin{equation}
\vect{N} = R\frac{\partial Z}{\partial\theta}\vect{e}_R 
+ \left( \frac{\partial R}{\partial\theta}\frac{\partial Z}{\partial\zeta} - \frac{\partial R}{\partial\zeta}\frac{\partial Z}{\partial\theta}\right)\vect{e}_\zeta
-R \frac{\partial R}{\partial\theta}\vect{e}_Z.
\end{equation}
We also have
\begin{equation}
\frac{\partial\vect{r}}{\partial R_{m,n}^c} =\cos(m\theta - n\zeta) \vect{e}_R,
\hspace{0.2in}
\frac{\partial\vect{r}}{\partial Z_{m,n}^c} =\cos(m\theta - n\zeta) \vect{e}_Z.
\end{equation}
The same expressions with $\cos \to \sin$ hold for $\partial\vect{r}/\partial R_{m,n}^s$
and $\partial\vect{r}/\partial Z_{m,n}^s$.
From (\ref{eq:Sdef3D}) then
\begin{align}
\frac{\partial f}{\partial R_{m,n}^s} &= \int_0^{2\pi}d\theta \int_0^{2\pi}d\zeta \sin(m\theta-n\zeta) S R \frac{\partial Z}{\partial\theta}, \label{eq:Rmns} \\
\frac{\partial f}{\partial R_{m,n}^c} &= \int_0^{2\pi}d\theta \int_0^{2\pi}d\zeta \cos(m\theta-n\zeta) S R \frac{\partial Z}{\partial\theta}, \label{eq:Rmnc} \\
\frac{\partial f}{\partial Z_{m,n}^s} &= -\int_0^{2\pi}d\theta \int_0^{2\pi}d\zeta \sin(m\theta-n\zeta) S R \frac{\partial R}{\partial\theta}, \label{eq:Zmns} \\
\frac{\partial f}{\partial Z_{m,n}^c} &= -\int_0^{2\pi}d\theta \int_0^{2\pi}d\zeta \cos(m\theta-n\zeta) S R \frac{\partial R}{\partial\theta}. \label{eq:Zmnc} 
\end{align}

We can now give the `only if' half of the proofs. Suppose the surface shape and $S$ are stellarator-symmetric. This symmetry means that under the transformation
$(\theta,\zeta) \to (-\theta,-\zeta)$, the quantities $R$, $\partial Z/\partial\theta$, and $S$ are even, and $Z$ and $\partial R/\partial\theta$ are odd.
The integrands in (\ref{eq:Rmns}) and (\ref{eq:Zmnc}) are odd, so the integrals vanish. Therefore $\partial f/\partial R_{m,n}^s=0$ and $\partial f/\partial Z_{m,n}^c=0$ 
for all $m$ and $n$.

Suppose the surface shape and $S$ are $n_{fp}$-symmetric, meaning $R$, $Z$, and $S$ only include Fourier mode numbers in $\zeta$ that are integer multiples of $n_{fp}$.
Then in (\ref{eq:Rmns})-(\ref{eq:Zmnc}), $S R \, \partial Z/\partial\theta$ and $S R \, \partial R/\partial\theta$ have the same property.
Therefore, if $n$ is not a multiple of $n_{fp}$, the integrals in (\ref{eq:Rmns})-(\ref{eq:Zmnc}) all vanish. This concludes the `only if' half of the proofs.

Now we give the `if' half of the proofs. Suppose the surface shape is stellarator-symmetric, so $R$ and $\partial Z/\partial\theta$ are even and $Z$ and $\partial R/\partial\theta$ are odd under $(\theta,\zeta) \to (-\theta,-\zeta)$, but we do not yet know the parity of $S$. Suppose further that $\partial f/\partial R_{m,n}^s=0$ and $\partial f/\partial Z_{m,n}^c=0$ for all $m$ and $n$. Substituting a Fourier expansion of $S R \, \partial Z/\partial\theta$ into (\ref{eq:Rmns}), one finds the 
Fourier components of the odd part of $S R \, \partial Z/\partial\theta$ all vanish,
hence the odd part of $S R \, \partial Z/\partial\theta$ must vanish everywhere. Since $R$ is positive, it follows that the odd part of $S$ can be nonzero only where $\partial Z/\partial\theta$ vanishes. A similar argument applied to (\ref{eq:Zmnc})
shows that the odd part of $S$ can be nonzero only where $\partial R/\partial\theta$ vanishes. Except in the uninteresting case that the coordinate system is singular, 
at least one of $\partial R/\partial\theta$ or $\partial Z/\partial\theta$ is nonzero everywhere, so the odd part of $S$ must vanish, i.e. $S$ is stellarator-symmetric. 

Finally, suppose the surface shape is $n_{fp}$-symmetric, so $R$ and $Z$ only include Fourier mode numbers in $\zeta$ that are integer multiples of $n_{fp}$,
but we do not assume this symmetry of $S$. Suppose further that $\partial f/\partial R_{m,n}^s=0$, $\partial f/\partial R_{m,n}^c=0$, $\partial f/\partial Z_{m,n}^s=0$, and $\partial f/\partial Z_{m,n}^c=0$ for all $n$ that are not integer multiples of $n_{fp}$.
Substituting a Fourier expansion of $S R \, \partial Z/\partial\theta$ into (\ref{eq:Rmns})-(\ref{eq:Rmnc}), one finds the Fourier components of
$S R \, \partial Z/\partial\theta$ all vanish when $n$ is not a multiple of $n_{fp}$, so $S R \, \partial Z/\partial\theta$ is $n_{fp}$-symmetric.
As division by an $n_{fp}$-symmetric function preserves the symmetry, then $S$ is $n_{fp}$-symmetric, except perhaps where $\partial Z/\partial\theta=0$.
Repeating the same argument using (\ref{eq:Zmns})-(\ref{eq:Zmnc}), one concludes 
$S$ is $n_{fp}$-symmetric, except perhaps where $\partial R/\partial\theta=0$.
Since at least one of $\partial R/\partial\theta$ or $\partial Z/\partial\theta$ is nonzero everywhere for a nonsingular coordinate system,
$S$ is $n_{fp}$-symmetric everywhere.
This concludes the `if' half of the proofs.

\bibliography{sensitivity}

\end{document}